\documentclass{article}

\usepackage{csquotes}
\usepackage[english]{babel}

\usepackage[a4paper,top=2cm,bottom=2cm,left=3cm,right=3cm,marginparwidth=1.75cm]{geometry}

\usepackage{amsmath}
\usepackage{graphicx}
\usepackage[colorlinks=true, allcolors=blue]{hyperref}
\usepackage{ dsfont }
\usepackage{longtable}
\usepackage{array}
\usepackage{subcaption}
\usepackage{multicol}
\usepackage{setspace}
\usepackage{algorithm}
\usepackage{algorithmic}
\usepackage{tabularx}
\usepackage{comment}
\usepackage{multirow}
\usepackage{booktabs}
\usepackage{xcolor}
\usepackage{amssymb}
\usepackage{authblk}
\usepackage{lmodern}

\makeatletter
\patchcmd{\@maketitle}{\LARGE \@title}{\fontsize{16}{19.2}\selectfont\@title}{}{}
\makeatother

\title{Counting and Hardness-of-Finding Fixed Points \\ in Cellular Automata on Random Graphs}
\author{Cédric Koller, Freya Behrens, and Lenka Zdeborová}
\affil[]{\textit{École Polytechnique Fédérale de Lausanne (EPFL)}\\
\textit{Statistical Physics of Computation Laboratory}}

\date{}

\begin{document}
\maketitle

\begin{abstract}
We study the fixed points of outer-totalistic cellular automata on sparse random regular graphs. These can be seen as constraint satisfaction problems, where each variable must adhere to the same local constraint, which depends solely on its state and the total number of its neighbors in each possible state. Examples of this setting include classical problems such as independent sets or assortative/dissasortative partitions. We analyse the existence and number of fixed points in the large system limit using the cavity method, under both the replica symmetric (RS) and one-step replica symmetry breaking (1RSB) assumption. This method allows us to characterize the structure of the space of solutions, in particular, if the solutions are clustered and whether the clusters contain frozen variables. This last property is conjectured to be linked to the typical algorithmic hardness of the problem. We bring experimental evidence for this claim by studying the performance of the belief-propagation reinforcement algorithm, a message-passing-based solver for these constraint satisfaction problems.
\end{abstract}

\section*{Introduction}

Systems with seemingly simple local interactions have been known to possibly present complex and unexpected behavior at a macroscopic scale. Cellular automata (CAs) are a prime example of this phenomenon. Conway's game of life \cite{gardner_mathematical_1970}, probably the most famous cellular automaton, has for example been proven to be Turing complete \cite{rendell_turing_2015}. Self-reproducing cellular automata have been studied by von Neumann as early as the 1940s \cite{von_neumann_theory_1966}. Even elementary cellular automata have been shown to exhibit complex behaviors \cite{wolfram_universality_1984}, and some of them have been proven to be Turing complete \cite{cook_universality_2004}. CAs are usually defined on regular grids, which impose strong correlations between the cells. Following \cite{behrens_dynamical_2023}, we relax this structure by considering cellular automata evolving synchronously on random regular graphs. This will enable analysis of their behaviour in the large size limit. 

In this paper, we study the fixed points, or stationary configurations, of these cellular automata on random regular graphs. The stationary configurations can be seen as solutions to constraint satisfaction problems (CSPs) where each variable must adhere to the same local constraint. Constraint satisfaction problems, due to their analogies with disordered systems, have been extensively studied using methods from statistical physics. The existence and entropy of solutions can be obtained by studying the uniform measure over the solutions. The partition function of this measure, or its proxy the free entropy, then counts the number of solutions. The computation of the partition function is hard in general, but approximations of the free entropy can be computed using the replica symmetric cavity method \cite{Bethe, geffner_reverend_2022, yedidia_understanding_2003, Information_physics_and_computations}, which is particularly suited to sparse graphs. Refining the replica symmetric approach, one can introduce assumptions of replica symmetry breaking within the cavity formalism \cite{mezard_sk_1986,mezard_bethe_2001, mezard_cavity_2002}. These advancements lead to the development of new efficient algorithms \cite{mezard_random_2002, chavas_survey-propagation_2005}. The statistical physics approach also allows the study of phase transitions in the space of solutions, of which the clustering \cite{biroli_variational_2000, mezard_analytic_2002, mezard_landscape_2005} and the freezing \cite{zdeborova_phase_2007, zdeborova_constraint_2008} transitions are of particular interest due to their conjectured link to the typical computational algorithmic complexity of the problem \cite{zdeborova_statistical_2016, gamarnik_disordered_2022}.

In this paper, we systematically investigate the class of outer-totalistic CSPs on random \mbox{$3-$regular} graphs using the cavity method, with the aim to discover novel interesting problems and structural similarities between them. In Section \ref{sec:Background and notation}, we define the outer-totalistic CSPs and provide a brief overview of some outer-totalistic CSPs studied in the literature. In Section \ref{sec:Replica symmetric solution using belief propagation}, we derive the RS cavity fixed point equation, which we use to estimate the entropy of solutions. In Section~\ref{sec:Classification of outer-totalistic rules}, we detail the classification based on the entropy of solutions. Section~\ref{sec:One-step replica symmetry breaking} introduces the general one-step replica symmetry breaking formalism, including an overview of the different phases, and the 1RSB cavity equations. The presence of replica symmetry breaking and its relation to the structure of the space of solutions is discussed in Section~\ref{sec:Structure of the space of solutions}. Finally, the connection between the phase and the experimental computational hardness of the outer-totalistic CSPs is explored in Section~\ref{sec:Freezing and computational hardness}.

The code to generate the data and plot the figures can be found at \url{https://github.com/SPOC-group/fixed_points_graph_CA}.

\section{Background and Notation}\label{sec:Background and notation}

\paragraph{Outer-totalistic constraint satisfaction problems.}
Consider an undirected $d$-regular graph $G$ with $N$ nodes $i=\{1, \hdots,n\}$. We denote the set of edges $E(G)$ and the edge between node $i$ and $j$ as $ij$. Each node $i$ is in a state $s_i\in S$ from a discrete alphabet $S$. In this work $S=\{0,1\}$, and we denote node $i$ as empty if $s_i=0$ and occupied if $s_i=1$. We refer to $\mathbf{s}=(s_1,...,s_N)$ as a \textit{configuration}. A configuration is said to be \textit{homogeneous} if all nodes are in the same state, i.e. $s_i=s_j\,\,\forall i,j=1,\hdots,N$.

A configuration is said to be a \textit{solution} of a CSP if all its nodes respect all the given constraints. In our case, every node of the graph must respect the same local constraint. We call this constraint a \textit{rule}. Let us denote as $f(s_i, \{s_j\}_{j\in \partial i})$ the indicator function that is $1$ if node $i$ with value $s_i$ respects the rule when its direct neighbors $\partial i$ take values $\{s_j\}_{j\in \partial i}$, and $0$ if it does not. We study \textit{outer-totalistic} rules, rules that depend only on the value of a node and the total number of occupied nearest neighbors. In the case where $S=\{0,1\}$, we have that $f(s_i, \{s_j\}_{j\in \partial i})=f(s_i, \sum_{j\in\partial i}s_j)$, but we will keep the first notation for generality.

To denote an outer-totalistic rule, we extend the notation introduced by \cite{marr_outer-totalistic_2009}. This notation consists of an ordered list of $d+1$ symbols, indicating what value a node can take to satisfy the constraint for $0,1,\hdots,d$ occupied neighbors. The symbols are
\begin{itemize}
    \item [\texttt{0}]: the node must be empty to satisfy the constraint,
    \item [\texttt{1}]: the node must be occupied to satisfy the constraint,
    \item [\texttt{+}]: the constraint is satisfied regardless of the value of the node, 
    \item[\texttt{-}]: the constraint cannot be satisfied. 
\end{itemize}  As an example, consider rule \texttt{-+10}: the constraint is not satisfied if a node has $0$ occupied neighbors, the node can take any value for $1$ occupied neighbor, it must be occupied for $2$ occupied neighbors, and it must be empty if it has $3$ occupied neighbors. Note that inverting the empty and occupied states gives rise to equivalent rules. To avoid studying twice the same problem, we remove this equivalence with the procedure explained in Appendix~\ref{appendix:Non-equivalent outer-totalistic rules}. For example, we have only $136$ non-equivalent rules out of $4^4=256$ rules for $d=3$. This notation was originally used to describe the dynamics of outer-totalistic cellular automata, but we adopt it here to describe static CSPs. While we use the cavity method to study the static case, note that it can also be used to study dynamical systems, for instance in \cite{behrens_backtracking_2023} where the same rule notation as here is used.

Note that this notation introduces a hierarchy between the rules: starting from rule \texttt{++++} for which every configuration is a solution, one can add constraints, first turning \texttt{+} into \texttt{0} or \texttt{1} and then into \texttt{-}. This gives a simple criterion when comparing the number of solutions of a given rule: all the rules with more constraints should have at most the same number of solutions.

\paragraph{Number of stationary configurations.}
Our aim is to count the number of solutions to a given outer-totalistic rule. The uniform probability measure over solutions of an outer-totalistic rule on a given $d$-regular graph $G$ is
\begin{equation}\label{eq:probability_measure_stationary_configuration}
    P_G(\mathbf{s})=\frac{\prod_{i=1}^N f(s_i, \{s_j\}_{j\in\partial i})}{Z_G},
\end{equation}
where $f(s_i, \{s_j\}_{j\in\partial i})$ indicates whether the rule is satisfied by taking the value $1$, and taking the value $0$ otherwise. $Z_G$ is the normalization constant (i.e. the partition function) and is given by
\begin{equation}\label{eq:partition_function_microcanonical}
    Z_G=\sum_{\mathbf{s}\in\mathcal{C}}\prod_{i=1}^N f(s_i, \{s_j\}_{j\in\partial i})
\end{equation}
where $\sum_{\mathbf{s}\in\mathcal{C}}$ indicates a sum over all possible configurations $\mathcal{C}=\{0,1\}^N$. Then, $Z$ is the number of stationary configurations. We define the \textit{free entropy density} $\phi_G$ on a given $d$-regular graph $G$ as
\begin{equation}\label{eq:free entropy density given graph}
    \phi_G = \frac{\log{Z_G}}{N}.
\end{equation}

These quantities can be analyzed on $d$-regular graphs in the large size limit. The cavity method (see Section~\ref{sec:Replica symmetric solution using belief propagation}) allows us to compute the \textit{quenched free entropy density} $\phi$ defined as
\begin{equation}\label{eq:quenched free entropy} 
    \phi=\lim_{N\rightarrow \infty} \frac{1}{N}\mathbb{E}_G [\log Z_G]
\end{equation}
where $\mathbb{E}_G$ denotes the average over the ensemble of random $d$-regular graphs of $N$ nodes. In statistical physics, it is standard that the quenched free entropy density, as well as other averaged quantities of interest, are self-averaging, i.e. concentrate around their mean. In particular, this means that the free entropy density~\eqref{eq:free entropy density given graph} for a typical large $d$-regular graph $G$ is equal to the quenched free entropy density~\eqref{eq:quenched free entropy} with high probability. It is instructive to also define the annealed free entropy density $\phi_{\text{ann}}$
\begin{equation}\label{eq:annealed free entropy}
    \phi_{\text{ann}}=\lim_{N\rightarrow \infty} \frac{1}{N}\log \left(\mathbb{E}_G[ Z_G]\right)
\end{equation}
that provides an upper bound to the quenched free entropy density.

\paragraph{Relation to other constraint satisfaction problems.}
Multiple well-studied CSPs are encompassed within the introduced rule notation. We present a non-exhaustive list of them, along with relevant literature that studied them using statistical physics approaches, in particular the cavity method.

The independent set or its complement, the vertex cover, was studied using the cavity method in \cite{weigt2000number,barbier_hard-core_2013}. The corresponding rules are of the form \texttt{+0$\hdots$0}. To study the maximum independent set, a Lagrange multiplier $\mu$ can be introduced and taken to infinity. 

The outer-totalistic CSPs are related to the lattice glass model proposed in \cite{biroli2001lattice} and further studied by Rivoire et al. \cite{rivoire_glass_2003}. In this model, occupied nodes can have at most $k$ occupied neighbors. This corresponds to rule \texttt{+$\hdots$+0$\hdots$0}. They obtained an analytic expression for the densest packing ($\mu\rightarrow \infty$) within the 1RSB approximation.

Another problem that can be recast in the language of outer-totalistic CSPs is the maximal independent set problem (rule \texttt{10$\hdots$0}), studied with the cavity method in \cite{dall2009statistical}, or the dominating set problem (rule \texttt{1+$\hdots$+}),
 studied with the cavity method in \cite{zhao2015statistical}. 

The assortative/dissasortative partition problems were studied in \cite{behrens_disassortative_2022}. The corresponding rules are of the form \texttt{0$\hdots$0$\pm$$\hdots$$\pm$1$\hdots$1} and \texttt{1$\hdots$1$\pm$$\hdots$$\pm$0$\hdots$0}. These problems were studied for balanced partitions, where the number of occupied and empty nodes are the same. This balance can be imposed by choosing an appropriate value of the parameter $\mu$ or by forcing the $0\leftrightarrow 1$ symmetry of the messages.

 Occupation problems on hyper-graphs were studied in \cite{zdeborova_constraint_2008}, and interestingly the averaged entropy of their solutions correspond to outer-totalistic rules composed only of \texttt{+} and \texttt{-}. More specifically, the free entropy obtained in the replica symmetric case is the same even if the underlying graphical model is not, and there is no clear mapping from the solutions of one to the solutions of the other on a single instance. The BP equation in present outer-totalistic CSPs seem to reduce to the BP equation in occupation problems from \cite{zdeborova_constraint_2008}, but we were only able to show this by assuming a structure in our messages that we do not know how to justify. The correspondence was verified numerically for $d=3,4$. See Appendix~\ref{appendix:occupation problems} for a more detailed discussion.

\section{Replica Symmetric Solution using Belief Propagation}\label{sec:Replica symmetric solution using belief propagation}

 The free entropy density for a given graph~\eqref{eq:free entropy density given graph} and the quenched free entropy density~\eqref{eq:quenched free entropy} are in general challenging to compute. Both can be estimated on large random graphs using the replica symmetric cavity method \cite{yedidia_understanding_2003, Information_physics_and_computations}. The approximation on a given graph $G$ is known as the \textit{Bethe-Peierls approximation}, and yields the so-called \textit{Bethe free entropy}. Averaging the Bethe free entropy over the ensemble of $d$-regular graphs gives the \textit{replica symmetric approximation} of the quenched free entropy density~\eqref{eq:quenched free entropy}. If this approximation is exact, we say that the problem is \textit{replica symmetric} (RS). The cavity method yields a fixed-point equation, which can be iterated as an algorithm. This algorithm is known as \textit{belief propagation} (BP), \textit{message passing} or \mbox{\textit{sum-product}}. We use the general method of casting the problem into a graphical model to obtain the BP equations.

\paragraph{Graphical model.}

\begin{figure}[ht]
  \centering
  \begin{subfigure}{0.45\textwidth}
    \centering
    \includegraphics[width=\textwidth]{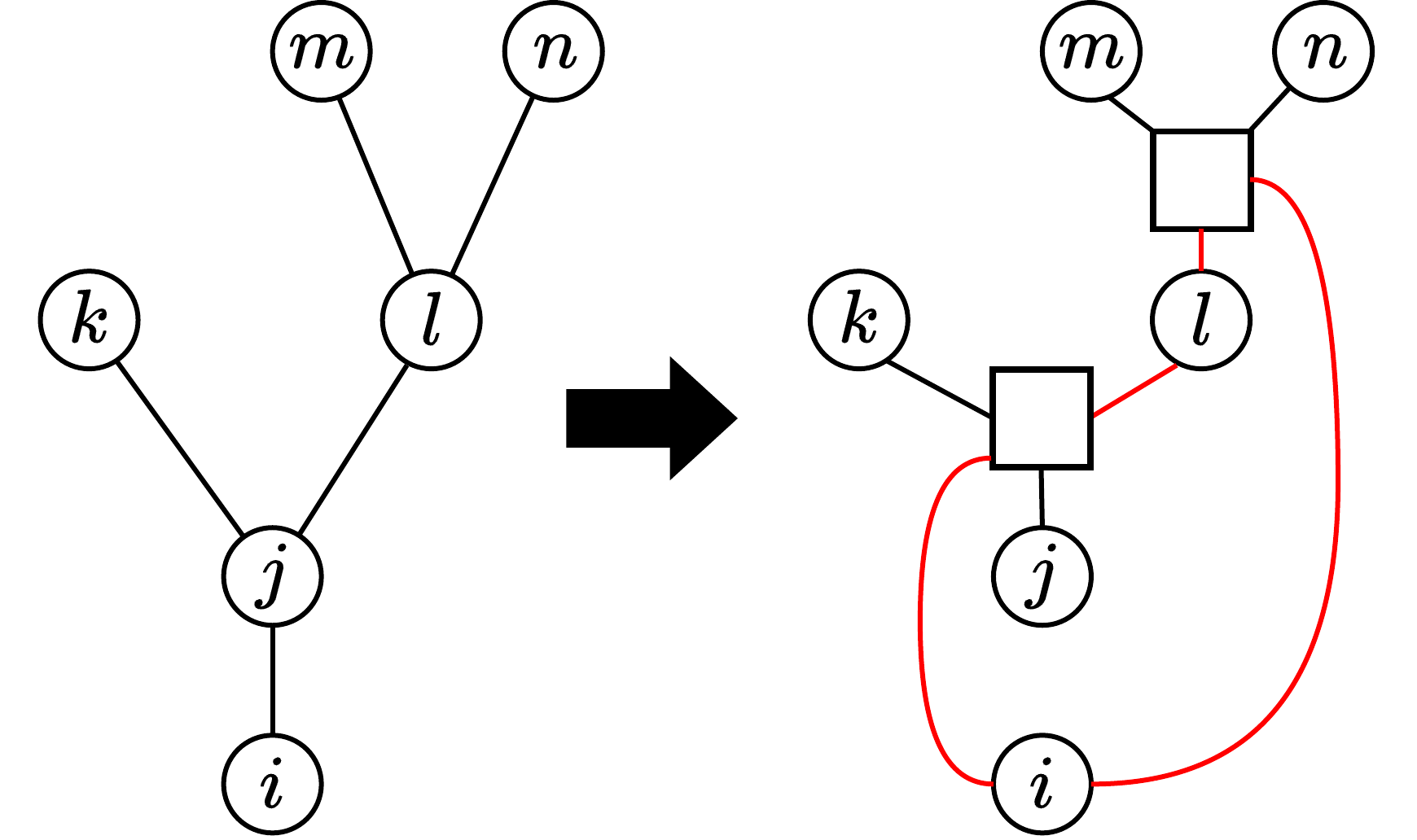} 
  \end{subfigure}
  \hfill
  \begin{subfigure}{0.45\textwidth}
    \centering
    \includegraphics[width=\textwidth]{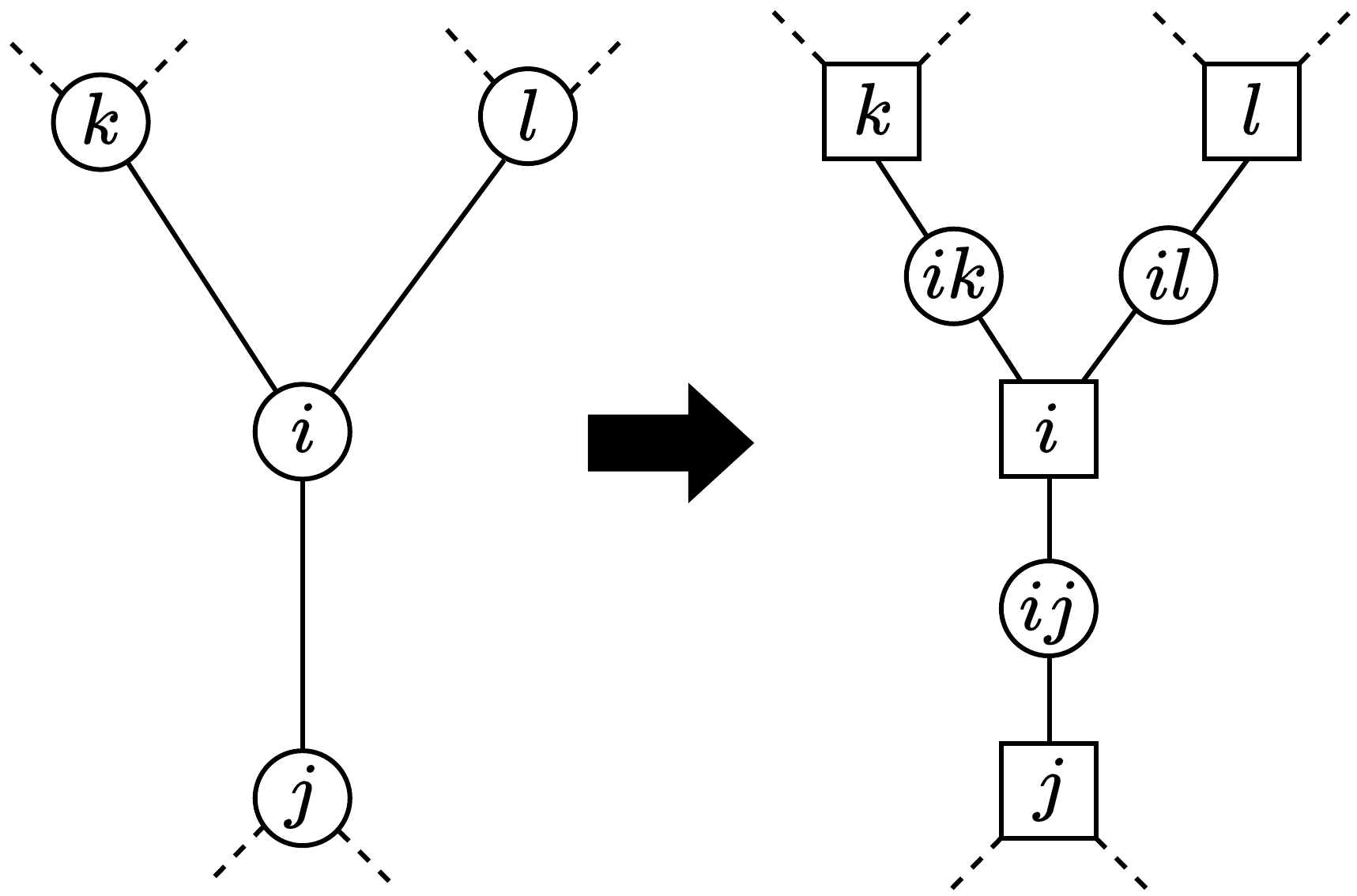} 
  \end{subfigure}
  \caption{Left: original graph and naive factor graph. The edges forming a short loop are drawn in red. Right: original graph and factor graph. Notice that the nodes become the factor nodes and the edges the variable nodes.}
  \label{fig:factor_graphs}
\end{figure}

A \textit{graphical model} is the representation of a factorized probability distribution, such as eq. \eqref{eq:probability_measure_stationary_configuration} on a \textit{factor graph}. A factor graph is a graph with two types of nodes: \textit{variable nodes}, one for each variable, and \textit{factor nodes}, one for each interaction (or constraint in our case). Variable nodes are usually drawn as circles and factor nodes as squares. An edge is present between a variable node and a factor node if its associated constraint depends on the value of the variable. The factor graph is thus bipartite: edges exist only between variable nodes and factor nodes. 

The naive way of constructing the graphical model of the probability distribution \eqref{eq:probability_measure_stationary_configuration} is to have one variable for each state $s_i$, and a factor node that checks if the constraint on $s_i$ is satisfied. We say that the graph $G$ on which the problem is set is the \textit{original graph} (or simply the graph when clear from the context). An example of an original graph and associated factor graph is illustrated in Figure~\ref{fig:factor_graphs} (left). We notice that, even if the original graph is a tree, the factor graph contains short loops (in red). This can be problematic, because the belief propagation algorithm presented below is heuristically known to work better on factor graphs with large loops \cite{notes}. Therefore, we use an alternative graphical model.

Consider the variables to be on the edges and the factor nodes to be on the nodes of the original graph. This means that the variables are tuples $s_i,s_j$. We denote the variables nodes as $ij$ and the factor nodes as $i$ (since they are equivalent to the nodes of the original graph). The variable $s_i, s_j$ take value in $S\times S$. For $S=\{0,1\}$, this gives 4 different possible values. By construction, the factor nodes have an edge towards each variable needed to check the constraint. An example of the original graph and obtained factor graph is presented in Figure~\ref{fig:factor_graphs} (right). Notice that the factor graph has exactly the same structure as the original graph. In particular, the loops are the same as in the original graph. Similar constructions are done e.g. in \cite{dall2009statistical, behrens_disassortative_2022, behrens_backtracking_2023}.

\paragraph{Belief propagation on a given graph.}
From the graphical model, we can deduce a fixed point equation named the \textit{replica symmetric cavity} or \textit{belief propagation} (BP) equation. The equation is derived in detail in Appendix~\ref{appendix:derivation BP on graphs} for the case of trees, where the Bethe-Peierls approximation is exact and yields the true free entropy density~\eqref{eq:free_entropy_graph}. A broader condition for BP to be exact is the independence of the incoming messages for each variable node. This is evidently the case if the factor graph is a tree. However, one also heuristically expects locally tree-like factor graphs, such as typical $d$-regular graphs for large $N$ and constant $d$, to give good results if the correlations decay fast enough in the different loops. This motivates the construction of a graphical model with no short loops. The equation derived for tree graphs can be used heuristically on general graphs and is given by
\begin{equation}\label{eq:BP_equations_graph}
    \psi^{i\rightarrow j}_{s_i, s_j}=
    \frac{1}{Z^{i\rightarrow j}}\sum_{\{s_k\}_{k\in \partial i \setminus j}} f(s_i, s_j, \{s_k\}_{k\in\partial i\setminus j}) \prod_{k\in \partial i \setminus j} \psi_{s_k, s_i}^{k\rightarrow i}.
\end{equation}
where we write $\{s_k\}_{k\in\partial i}$ as $s_j, \{s_k\}_{k\in\partial i\setminus j}$ to emphasise that $s_j$ is fixed by $\psi^{i\rightarrow j}_{s_i, s_j}$. We denote a configuration of the neighbors of node $i$ with node $j$ removed as $\{s_k\}_{k\in\partial i\setminus j}$, and the sum is over all these possible configurations. $\psi^{i\rightarrow j}_{s_i, s_j}$ is called the \textit{message} from node $i$ to $j$, and can be used to compute the free entropy density, and thus also other quantities of interest. $Z^{i\rightarrow j}$ is the normalization such that $\sum_{s_i}\sum_{s_j} \psi^{i\rightarrow j}_{s_i, s_j}=1$. The Bethe free entropy density $\phi_{\text{Bethe}}$ for a given graph $G$, also derived in Appendix~\ref{appendix:derivation BP on graphs}, is given by
\begin{equation}\label{eq:free_entropy_graph}
    \phi_{\text{Bethe}}=\frac{1}{N}\sum_{i=1}^N \log Z^i -\frac{1}{N}\sum_{ij\in E(G)}\log Z^{ij}
\end{equation}
with
\begin{equation}
    Z^i=\sum_{s_i}\sum_{\{s_k\}_{k\in\partial i}} f(s_i, \{s_k\}_{k\in\partial i}) \prod_{k\in\partial i}\psi_{s_k,s_i}^{k\rightarrow i},
\end{equation}
\begin{equation}
    Z^{ij}=\sum_{s_i}\sum_{s_j}\psi_{s_{(ij)}}^{i\rightarrow j} \psi_{s_j,s_i}^{j\rightarrow i}.
\end{equation}

\paragraph{Belief propagation on large random $d$-regular graphs.}
In this work, we study the outer-totalistic CSPs on large random $d$-regular graphs. Under the assumption that all the nodes are locally equivalent, which is motivated by the tree-like structure of random regular graphs, the fixed point equation in the $N\rightarrow \infty$ limit reduces to 4 self-consistent equations:
\begin{equation}\label{eq:BP}
      \psi_{s_i, s_j}=
    \frac{1}{Z_\psi}\sum_{\{s_k\}_{k=1}^{d-1}} f(s_i, s_j, \{s_k\}_{k=1}^{d-1}) \prod_{k=1}^{d-1} \psi_{s_k, s_i}.
\end{equation}
Recall that $f(s_i, s_j, \{s_k\}_{k=1}^{d-1})$ is $1$ if the constraint is satisfied and $0$ otherwise. $Z_\psi$ is the normalization and is given by
\begin{equation}
    Z_\psi=\sum_{s_i}\sum_{s_j}\sum_{\{s_k\}_{k=1}^{d-1}} f(s_i, s_j, \{s_k\}_{k=1}^{d-1}) \prod_{k=1}^{d-1} \psi_{s_k, s_i}.
\end{equation}
The iteration scheme to obtain a fixed point is presented in Appendix~\ref{appendix:Belief propagation iterations on d-regular graphs}. As before, the messages allow us to compute what we call the \textit{replica symmetric free entropy density} $\phi_{\rm RS}$:
\begin{equation}\label{eq:RS_free_entropy}
\phi_{\rm RS}=\log \left(Z_\text{n}\right)
-\frac{d}{2}\log\left(Z_\text{e}\right),
\end{equation}
where 
\begin{equation}\label{eq:Z_n}
    Z_\text{n}=\sum_{s_i, \{s_k\}_{k=1}^d} f(s_i,\{s_k\}_{k=1}^d) \prod_{k=1}^{d}\psi_{s_k, s_i},
\end{equation}
\begin{equation}\label{eq:Z_e}
    Z_\text{e}=\sum_{s_i, s_j}\psi_{s_i, s_j}\psi_{s_j, s_i}.
\end{equation}
Note that in this setting the graph is no longer specified. Moreover, in the specific case of regular graphs, it can be shown that the replica symmetric free entropy is equal to the annealed free entropy~\eqref{eq:annealed free entropy} \cite{mora_geometrie_2007} so that we have $\phi_{\text{ann}}=\phi_{\rm RS}$. In the following sections we will investigate for which rules we also have equality  $\phi_{\rm RS}=\phi$ with the quenched free energy defined in \eqref{eq:quenched free entropy}.

\paragraph{Stability and validity of the RS solution.}
A necessary (but not sufficient) condition for the validity of the belief propagation and the resulting free entropy is the so-called stability \cite{almeida_stability_1978,zdeborova_statistical_2008}. Stability is equivalent to the convergence of the BP equations on a single large graph. This can be verified numerically using a population dynamics algorithm. The algorithm is presented in Appendix \ref{appendix: population dynamics}. We investigate the stability of each rule by initializing the population dynamics with the BP messages with added noise. We use Gaussian noise with standard deviation $10^{-6}$, $\psi_{s_i, s_j}^{\rm noisy}=\max(\psi_{s_i, s_j}+\eta,0), \,\eta\sim\mathcal{N}(0, 10^{-12})$, which we then normalize. If the population dynamics goes back to the BP fixed point, the fixed point is stable \cite{zdeborova_statistical_2008}.

If the fixed point is not stable, then the replica symmetric assumption is incorrect and the problem presents replica symmetry breaking. In that case, we analyse the problem within the one-step replica symmetry breaking (1RSB) assumption, which is presented in Section \ref{sec:One-step replica symmetry breaking}. Note that the reciprocal is not necessarily true: the replica symmetric assumption can be incorrect even if the RS fixed point is stable. 

We find that the RS solution is stable for all rules with existing solutions ($\phi\geq 0$), except for two rules: \texttt{+-00} and \texttt{+-0+}. As will be seen in Section \ref{sec:Structure of the space of solutions}, these rules indeed present replica symmetry breaking. More generally, the validity of the RS solution can also be checked by doing the 1RSB calculation: if only the trivial 1RSB solutions exist (i.e. the only 1RSB solution is the replica symmetric one), then the RS solution is correct. This was verified for all rules marked `RS' in Tables \ref{tbl:classification_with_homogeneous} and \ref{tbl:classification_without_homogeneous}.

\section{Classification of Outer-totalistic Rules for \texorpdfstring{$d=3$}{LG}}\label{sec:Classification of outer-totalistic rules}
We classify all possible non-equivalent outer totalistic rules depending on the number of solutions. To compute the RS fixed point, we start the BP iteration (see Appendix~\ref{appendix:Belief propagation iterations on d-regular graphs}) with $10^6$ different initializations $\psi_{s_i,s_j}=0,0.01, 0.02, ..., 1 \,\forall \,s_i, s_j \in S$. It is possible that multiple fixed points exist, and that they are found numerically depending on the initialization. In that case, the fixed point with the highest free entropy is kept. In the following section, we present a classification based on the number and type of solutions.

The first clear distinction is whether a rule admits homogeneous solutions or not. The formula of the rule clearly indicates if the rule admits a homogeneous solution. In this case, there are three possible scenarios:
\begin{itemize}
    \item \textbf{Only homogeneous solution:} The problem only admits homogeneous solutions. This means that $\phi=0$ and that the fixed point messages are only compatible with homogeneous solutions.
    \item \textbf{Subexponentially many solutions:} Other solutions than the homogeneous ones exist, but their number is subexponential in $N$, i.e. $\phi=0$.
    \item \textbf{Exponentially many solutions:} An exponential number of solutions exist ($\phi>0$) in addition to the homogeneous solutions.
\end{itemize}
Table~\ref{tbl:homogeneous rules} and \ref{tbl:classification_with_homogeneous} list all the rules that have homogeneous solutions, Table~\ref{tbl:homogeneous rules} listing the rules that have \textit{only} homogeneous solutions. Table \ref{tbl:classification_without_homogeneous} presents all the rules that have no homogeneous solutions.

In the latter case, the following possibilities arise:
\begin{itemize}
    \item \textbf{Locally contradictory:} It is not possible to find a solution on a $d$-regular tree of depth 2 for any boundary condition.
    \item \textbf{No solution:} The rule is not locally contradictory but the free entropy density is negative.
    \item \textbf{Subexponentially many solutions:} The free entropy density is $0$, there exists a subexponential number of solutions.
    \item \textbf{Exponentially many solutions:} The free entropy density is strictly positive, there is an exponentially number solutions.
\end{itemize}
 For the rules that are locally contradictory, either the message is not normalisable (all the entries of the messages are $0$) or there is no convergence. In this case, we define the free entropy density to be $-\infty$ and the phase as unsatisfiable (UNSAT) as there are no solutions. The rules with no solutions but that are not locally contradictory admit a fixed point with negative free entropy density. Thus, the obtained fixed point allows us to distinguish between the rules that are locally contradictory and the rules that have no solutions but are not locally contradictory.

The classification in Tables \ref{tbl:homogeneous rules}, \ref{tbl:classification_with_homogeneous} and \ref{tbl:classification_without_homogeneous} are done using the replica symmetric free entropy, which is not necessarily the same as the true free entropy. To study the validity of the replica symmetric free entropy and correct it if necessary, we compute the free entropy within the one-step replica symmetry breaking framework which is presented below in Sections \ref{sec:One-step replica symmetry breaking} and \ref{sec:Structure of the space of solutions}. As we will see, for $d=3$, most rules are replica symmetric. The inclusion of one-step replica symmetry breaking does not change the classification.

\section{One-step Replica Symmetry Breaking}\label{sec:One-step replica symmetry breaking}
In this section, we give a brief overview of one-step replica symmetry breaking viewed from the lens of the cavity method. The method was introduced in \cite{mezard_bethe_2001, mezard_cavity_2002}, and for more details we refer the reader to \cite{Information_physics_and_computations, notes}.
\paragraph{Clustering.}
When the replica symmetric assumption holds, almost all stationary solutions are said to be in a \textit{cluster} or \textit{pure state} in configuration space. 
When the replica symmetric assumption is not correct, we say that there is a \textit{replica symmetry breaking}. If the configuration space is correctly described by decomposing the replica symmetric cluster into smaller clusters, we say that there is a \textit{one-step replica symmetry breaking} (1RSB). For a given graph $G$ there are two messages $\psi^{i\rightarrow j}$ and $\psi^{j\rightarrow i}$ for each edge $ij\in E(G)$. We identify each cluster by the fixed point $\tilde{\psi}=\{\psi^{i\rightarrow j}, \psi^{j\rightarrow i}\}_{ij\in E(G)}$. The probability measure $\nu$ over the cluster associated to a fixed point $\tilde\psi$ is
\begin{equation}\label{eq:probability measure cluster}
    \nu(\tilde\psi)=\frac{1}{Z_{\nu}}e^{Nm\phi_{\rm int}(\tilde\psi)}
\end{equation}
where $m$ is the \textit{Parisi parameter}, $Z_\nu$ is the normalization and $\phi_{\rm int}(\tilde\psi)$ is the Bethe free entropy density associated to the fixed point $\tilde\psi$. We will call $\phi_{\rm int}$ the \textit{internal free entropy density}.
The free entropy density associated with the measure $\nu$ is called the \textit{replicated free entropy} \mbox{$\Psi(m)=\log(Z_\nu)/N$}. The \textit{complexity} or \textit{configurational entropy density} $\Sigma$ is the entropy density at the level of the clusters, i.e. $\mathcal{N}_c (\phi)=e^{N\Sigma(\phi_{\rm int})}$ where $\mathcal{N}_c (\phi_{\rm int})$ is the number of cluster with internal entropy $\phi_{\rm int}$. Then we have that
\begin{equation}
    e^{N\Psi(m)}=\sum_{\phi_{\rm int}}e^{N(m\phi_{\rm int}+\Sigma(\phi_{\rm int}))}
\end{equation}
where the sum is over all the possible different internal free entropy densities. Applying the Laplace method we obtain 
\begin{equation}\label{eq:Laplace replicated free entropy density}
    \Psi(m)=m\phi_{\rm int}+\Sigma(\phi_{\rm int})
\end{equation}
where $\phi_{\rm int}$ extremizes $\Psi(m)$. Thus, $ \partial \Sigma/\partial \phi_{\rm int}=-m$ from the properties of the Legendre transform. To obtain the internal free entropy $\hat{\phi}_{\rm int}$ of the clusters containing the typical solutions (i.e. the solutions that dominate the measure~\ref{eq:probability_measure_stationary_configuration}), we maximize the total free entropy $\phi_{\rm int}+\Sigma (\phi_{\rm int})$ under the constraint $\Sigma(\phi_{\rm int})\geq 0$. Indeed, assuming that solutions exist, there must be at least one cluster. If we first ignore this constraint, the maximal total entropy is obtained for $m=1$. If the complexity is positive, we call such a clustered phase the \textit{dynamical one-step replica symmetry breaking phase} (d1RSB). In that case, there exists an exponential number of clusters with internal free entropy $\hat{\phi}_{\rm int}$. However, if the corresponding complexity is negative for $m=1$, then the total free entropy is maximized for the largest internal free entropy density $\hat{\phi'}_{\rm int}$ such that $\Sigma(\hat{\phi'}_{\rm int})\geq 0$. We write $\hat{\phi'}_{\rm int}=\phi_s$. In that case we are in the \textit{static one-step replica symmetry breaking phase} (s1RSB), there are subexponentially many clusters with internal free entropy $\phi_s$. Since the RS free entropy is equal to the annealed free entropy in the case of regular graphs \cite{mora_geometrie_2007}, the s1RSB phase indicates that the annealed free entropy is different from the quenched free entropy. If no $\hat{\phi'}_{\rm int}$ exist such that $\Sigma(\hat{\phi'}_{\rm int})\geq 0$, then the problem is said to be \textit{unsatisfiable} (note that the problem can of course also be unsatisfiable without replica symmetry breaking). Figure~\ref{fig:complexity_curve} (left) shows a sketch of the space of solutions for the different phases.

\paragraph{Freezing.} A variable $s_i$ is said to be \textit{frozen} if it takes the same value for every configuration within a cluster. If an extensive number of variables are frozen within a cluster, we say that the cluster is \textit{frozen}. Using the usual nomenclature (summarized for example in \cite{gabrie_phase_2017}), the \textit{rigidity transition} happens when frozen variables appear in typical solutions, and such a phase is called \textit{rigid}. The transition where all clusters become frozen is called the \textit{freezing transition}, and such a phase is called \textit{frozen}. If almost all the variables within a cluster are frozen, the cluster is said to be \textit{locked}. In that case the internal free entropy of the cluster is $0$.

\paragraph{1RSB cavity solution.} To estimate $Z_\nu$, we write similar equations as BP for the probability distribution $\nu$. Again assuming that the graph is $d$-regular and that the messages are locally equivalent we obtain 
\begin{equation}\label{eq:1RSB}
    P(\psi)=\frac{1}{Z_P}\int \prod_{k=1,...,d-1}dP(\psi^{(k)}) \left( \frac{Z_\text{n}}{Z_\text{e}}\right)^m \delta\left(\psi-\mathcal{F}(\{\psi^{(k)}\}_{k=1,...,d-1})\right).
\end{equation}
where $P$ is a continuous probability distribution over the messages. The integral is over all possible $d-1$ messages that are normalized. $\psi$ and $\psi^{(k)}$ indicate the 4-component message introduced in eq. \eqref{eq:BP}.  $\mathcal{F}\left(\{\psi^{(k)}\}_{k=1,...,d-1}\right)$ indicates the message $\psi$ obtained from eq.~\eqref{eq:BP} with the substitution $\psi_{s_k, s_i}=\psi^{(k)}_{s_k, s_i}$. $Z_P$ is the normalization and $Z_\text{n}$ and $Z_\text{e}$ are defined in eq.~\eqref{eq:Z_n} and \eqref{eq:Z_e}. For notational purposes we do not explicitly write the dependence of $Z_\text{n}$ and $Z_\text{e}$ on the message. Supposing that the 1RSB solution holds, the replicated free entropy is given by
\begin{equation}\label{eq:1RSB replicated free entropy density}
    \Psi_{\rm 1RSB}(m)=\log\left(\mathcal{Z}_n\right)-\frac{d}{2}\log\left(\mathcal{Z}_e\right)
\end{equation}
where
\begin{equation}\label{eq:cal Z_n}
    \mathcal{Z}_n=\int \prod_{k=1,...,d-1}dP(\psi^{(k)}) (Z_\text{n})^m,
\end{equation}
\begin{equation}\label{eq:cal Z_e}
    \mathcal{Z}_e=\int dP(\psi^{(1)}) dP(\psi^{(2)}) (Z_\text{e})^m.
\end{equation}
The internal entropy density can be obtained from
\begin{equation}\label{eq:internal_free_entropy_derivative}
 \phi_{\rm int}=\frac{\partial \Psi(m)}{\partial m}
 \end{equation}
which allows to compute the complexity using \eqref{eq:Laplace replicated free entropy density}. A detailed derivation of eq. \eqref{eq:1RSB}, \eqref{eq:1RSB replicated free entropy density} and \eqref{eq:internal_free_entropy_derivative} is given in Appendix~\ref{appendix: One-step replica symmetry breaking cavity equation}. Since the fixed point equation is now over a continuous distribution, we must usually approximate it numerically. For this, we use the population dynamics algorithm described in Appendix~\ref{appendix: population dynamics}. The fraction of frozen variables can be determined by counting the number of messages in the population that imply that a node has to take a fixed value.

\section{Structure of the Space of Solutions}\label{sec:Structure of the space of solutions}

\begin{figure}[ht]
  \begin{subfigure}{0.45\textwidth}
    \centering
    \includegraphics[width=\textwidth]{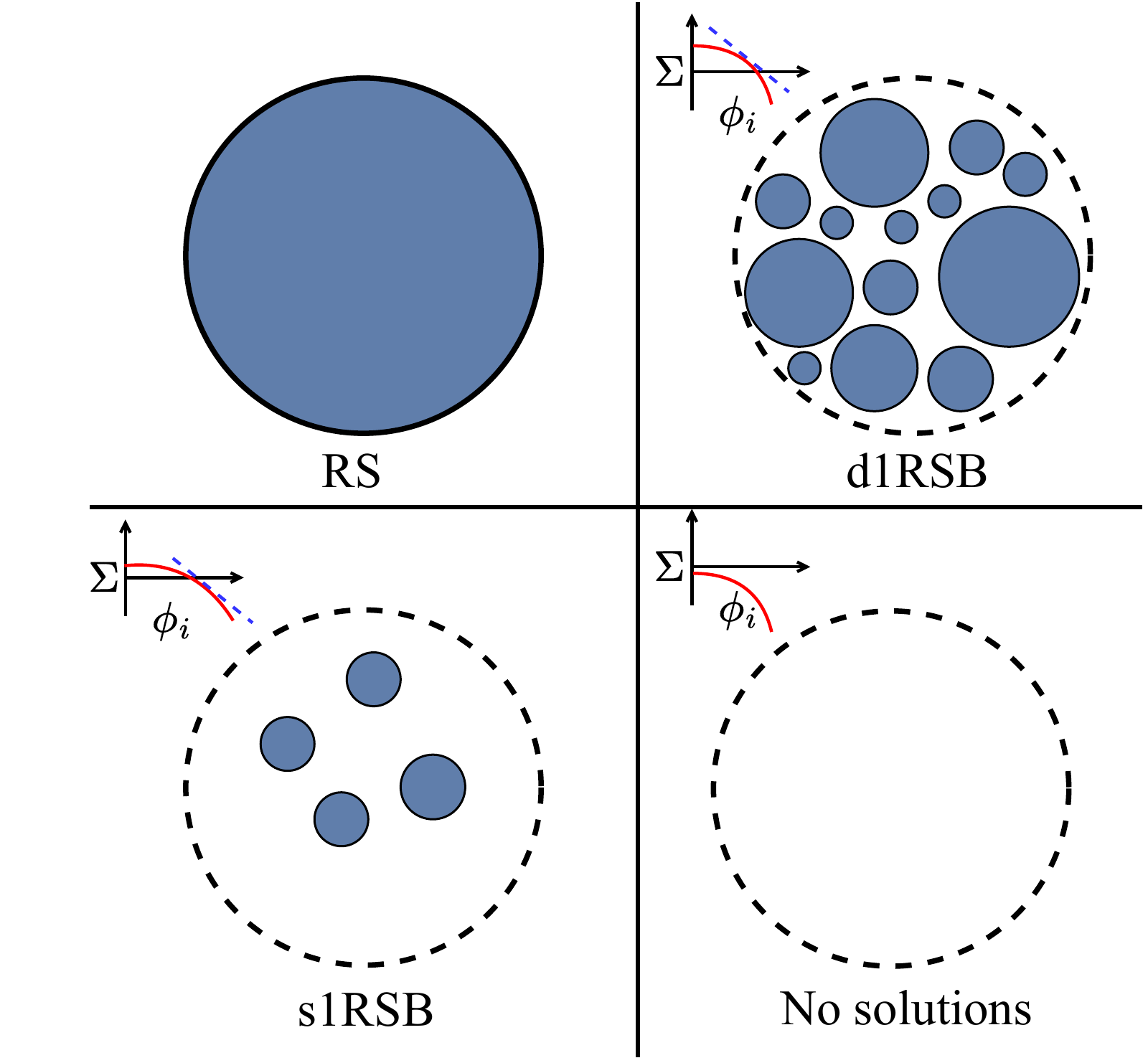} 
  \end{subfigure}
  \begin{subfigure}{0.45\textwidth}
    \centering
    \includegraphics[width=1.3\textwidth]{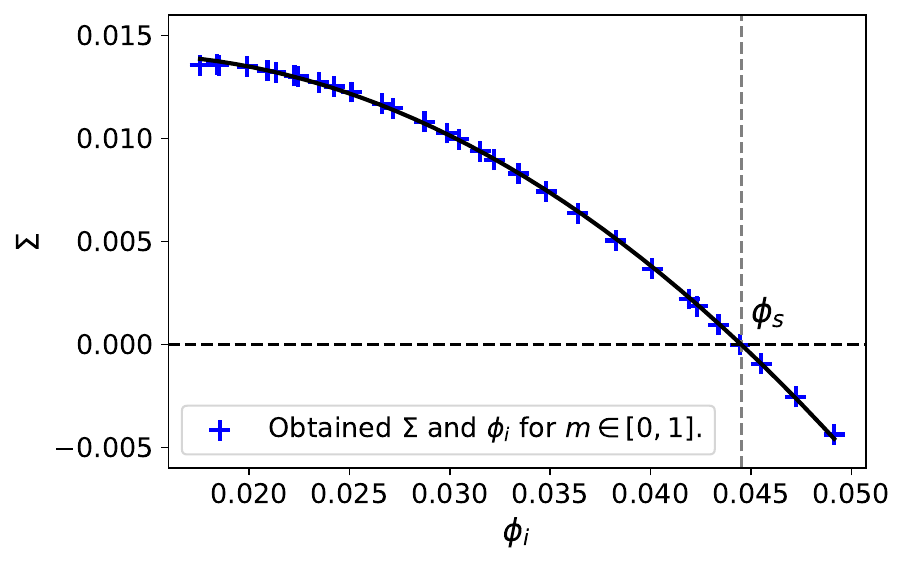}
  \end{subfigure}
  \caption{Left: Sketch of the space of configurations for the different phases. Blue indicate typical solutions. For the 1RSB phases the complexity vs internal entropy curves are sketch on the top left, the blue doted line indicates the tangent line where the slope of the complexity curve is $-1$, which is where $m=1$. Right: Complexity as function of the internal free entropy to estimate $\phi_s$ for the s1RSB rule \texttt{+-00}.
  }
  \label{fig:complexity_curve}
\end{figure}

\paragraph{Population dynamics and phases.}

As mentioned in Section \ref{sec:One-step replica symmetry breaking}, we estimate the solution to the 1RSB equation \eqref{eq:1RSB} using a population dynamics algorithm (see Appendix~\ref{appendix: population dynamics}). This also allows one to determine the phase. The population dynamics is conducted with a population of $M=10^7$ messages initialized on hard-fields (i.e. each message of the population dynamics is $1$ for one of the variables $(s_i, s_j)$, and $0$ on the others, which fixes the state of the nodes) proportionally to the obtained message from BP, e.g. if $\psi_{0,0}=0.3$ then we initialize $0.3M$ messages with $\psi_{0,0}=1$. The initialization on hard-fields is justified by the problem of reconstruction on trees. Indeed, the existence of non-trivial solutions is linked to the reconstruction on trees property \cite{mezard_reconstruction_2006}: consider a tree with a valid configuration, and remove all information about the state of the nodes except on the leaves. If the leaves contain some information about the value of the root, then the reconstruction on trees is possible. The initial population plays the role of the leaves and is thus initialized on hard-fields.

Another initialization of interest corresponds to the small noise reconstruction on trees. It is similar to the initialization on hard-fields, but with a small noise added on the messages. It can happen that both initializations do not give the same results, in particular that the hard-fields initialization yields a non-trivial solution whereas the small noise initialization converges to the RS fixed point. In that case, the solutions are `fake clustered' in the same sense as was proposed in \cite{zdeborova_constraint_2008} where a similar situation arises. In these cases, we posit that the RS solution in fact correctly describes the problem. An illustrative case of the phenomenon can be seen for rule \texttt{0010}. This rule admits as solutions all configurations that have non-intersecting loops. Thus, we can go from one solution to another by removing or adding one loop, which can be of subextensive length. This phenomenon seems to happen particularly for graphs with a small average degree. The `phase' of the rules were the hard-fields initialization has a non-trivial solution but the small noise initialization converges to the RS fixed point are noted RS*, and the complexity and internal free entropy are those obtained from the hard-field initialization.

The other phases are described in Sections \ref{sec:Classification of outer-totalistic rules}, \ref{sec:One-step replica symmetry breaking} and are
\begin{itemize}
    \item \textbf{UNSAT}: No RS fixed point or contradictory RS fixed point
    \item \textbf{RS}: Replica symmetric
    \item \textbf{d1RSB}: Dynamical one-step replica symmetry breaking
    \item \textbf{s1RSB}: Static one-step replica symmetry breaking
\end{itemize}
They are shown in both Tables \ref{tbl:classification_with_homogeneous} and \ref{tbl:classification_without_homogeneous}. The d1RSB phase is identified when the population dynamics solution at Parisi parameter $m=1$ is not trivial (i.e. not the same as the RS solution) and both the internal free entropy density and complexity are positive (see Figure~\ref{fig:complexity_curve} left for a sketch). The dynamical 1RSB phase holds the same entropy as the RS solution \cite{zdeborova_statistical_2008}. Frozen variables were found in all identified d1RSB rules. In brackets the standard deviation of the sampled observables at different iterations (see Appendix~\ref{appendix: population dynamics}) is indicated. This quantity is not indicated if it is smaller than $10^{-5}$.

\begin{table}[t]
\centering
\begin{tabular}{c}
\small
 Rules with only homogeneous solutions. \\
 \toprule
 \texttt{0000}, \texttt{000+}, \texttt{000-}, \texttt{0001}, \texttt{00-0}, \texttt{00-+}, \texttt{00--}, \texttt{00-1}, \texttt{0-00}, \texttt{0-0+}, \texttt{0-0-}, \texttt{0-01}, \texttt{0-+0}, \texttt{0-+-}, \texttt{0--0},\\
 \texttt{0--+}, \texttt{0---}, \texttt{0--1}, \texttt{0-10}, \texttt{0-1-}, \texttt{010-}, \texttt{01-0}, \texttt{01-+}, \texttt{01--}, \texttt{011-}, \texttt{+-0-}, \texttt{+--0}, \texttt{+--+}, \texttt{+---}, \texttt{+-10}, \\
 \texttt{+-1-}, \texttt{+1-0}, \texttt{+1--}, \texttt{+11-}. \\
 \bottomrule
\end{tabular}
\caption{All the non-equivalent rules with $d=3$ neighbors that only have homogeneous solutions.}
\label{tbl:homogeneous rules}
\end{table}

\begin{table}[ht]
\setlength{\tabcolsep}{4pt}
\renewcommand{\arraystretch}{1.2}
\centering
\tiny
\subfloat{
\begin{tabular}{|c|c|c|c|c|c|}
            \hline
            \multicolumn{6}{|c|}{Subexponentially many solutions} \\
            \hline
            \hline
            rule & phase & $\phi_{\rm RS}$ & $\phi_{\rm int}, m=1$ & $\Sigma, m=1$ & $\phi_{s}$ \\
            \hline
 \texttt{0101} & RS    & 0.0000 &        &            &  \\
 \texttt{+-+-} & d1RSB & 0.0000 & 0.0000 & 0.0000 &  \\
            \hline
            \hline
            \multicolumn{6}{c}{} \\
\hline
\multicolumn{6}{|c|}{Exponentially many solutions} \\
\hline
\hline
rule & phase & $\phi_{\rm RS}$ & $\phi_{\rm int}, m=1$ & $\Sigma, m=1$ & $\phi_{s}$ \\
\hline
 \texttt{00+0} & RS* & 0.2046          & 0.0000  & 0.2046(1) &        \\
 \texttt{00++} & RS             & 0.4133          &                     &                    &        \\
 \texttt{00+-} & RS* & 0.1925          &  0.0000 &  0.1925(2) &        \\
 \texttt{00+1} & RS             & 0.3793          &                     &                    &        \\
 \texttt{0010} & RS* & 0.1204          & 0.0000   & 0.1203(1) &        \\
 \texttt{001+} & RS             & 0.2833          &                     &                    &        \\
 \texttt{001-} & RS* & 0.1160          & 0.0000   & 0.1160(1) &        \\
 \texttt{0011} & RS             & 0.2350          &                     &                    &        \\
 \texttt{0+00} & RS             & 0.3671          &                     &                    &        \\
 \texttt{0+0+} & RS             & 0.3967          &                     &                    &        \\
 \texttt{0+0-} & RS             & 0.3484          &                     &                    &        \\
 \texttt{0+01} & RS             & 0.3740          &                     &                    &        \\
 \texttt{0++0} & RS             & 0.5548          &                     &                    &        \\
 \texttt{0+++} & RS             & 0.6350          &                     &                    &        \\
 \texttt{0++-} & RS             & 0.5134          &                     &                    &        \\
 \texttt{0++1} & RS             & 0.5829          &                     &                    &        \\
 \texttt{0+-0} & RS             & 0.2443          &                     &                    &        \\
 \texttt{0+-+} & RS             & 0.2601          &                     &                    &        \\
 \texttt{0+--} & RS             & 0.2359          &                     &                    &        \\
 \texttt{0+-1} & RS             & 0.2497          &                     &                    &        \\
 \texttt{0+10} & RS             & 0.3603          &                     &                    &        \\
 \texttt{0+1+} & RS             & 0.4265          &                     &                    &        \\
 \texttt{0+1-} & RS             & 0.3348          &                     &                    &        \\
 \texttt{0-++} & RS             & 0.3456          &                     &                    &        \\
 \texttt{0-1+} & RS             & 0.2833          &                     &                    &        \\
 \texttt{0100} & d1RSB          & 0.0416          & 0.0107(2)           & 0.0309(2)          &        \\
 \texttt{010+} & d1RSB          & 0.1303          & 0.0071(2)           & 0.1232(2)          &        \\
 \texttt{01+0} & RS             & 0.3715          &                     &                    &       
\end{tabular}}
\quad
\subfloat{
\begin{tabular}{|c|c|c|c|c|c|}
 \texttt{01++} & RS             & 0.5219          &                     &                    &        \\
 \texttt{01+-} & RS             & 0.2107          &                     &                    &        \\
 \texttt{0110} & RS             & 0.1893          &                     &                    &        \\
 \texttt{011+} & RS             & 0.4131          &                     &                    &        \\
 \texttt{+000} & RS             & 0.4354          &                     &                    &        \\
 \texttt{+00+} & RS             & 0.4354          &                     &                    &        \\
 \texttt{+00-} & RS             & 0.4131          &                     &                    &        \\
 \texttt{+0+0} & RS             & 0.4354          &                     &                    &        \\
 \texttt{+0++} & RS             & 0.4616          &                     &                    &        \\
 \texttt{+0+-} & RS             & 0.4131          &                     &                    &        \\
 \texttt{+0-0} & RS             & 0.2938          &                     &                    &        \\
 \texttt{+0-+} & RS             & 0.2938          &                     &                    &        \\
 \texttt{+0--} & RS             & 0.2833          &                     &                    &        \\
 \texttt{+010} & RS             & 0.2938          &                     &                    &        \\
 \texttt{+01+} & RS             & 0.2964          &                     &                    &        \\
 \texttt{+01-} & RS             & 0.2833          &                     &                    &        \\
 \texttt{++00} & RS             & 0.5513          &                     &                    &        \\
 \texttt{++0+} & RS             & 0.5605          &                     &                    &        \\
 \texttt{++0-} & RS             & 0.5143          &                     &                    &        \\
 \texttt{+++0} & RS             & 0.6489          &                     &                    &        \\
 \texttt{++++} & RS             & 0.6931          &                     &                    &        \\
 \texttt{+++-} & RS             & 0.5975          &                     &                    &        \\
 \texttt{++-0} & RS             & 0.3610          &                     &                    &        \\
 \texttt{++-+} & RS             & 0.3662          &                     &                    &        \\
 \texttt{++--} & RS             & 0.3415          &                     &                    &        \\
 \texttt{++10} & RS             & 0.4233          &                     &                    &        \\
 \texttt{++1-} & RS             & 0.3900          &                     &                    &        \\
 \texttt{+-00} & s1RSB          & 0.0448          & 0.0491(2)           & -0.0044(1)         & 0.0445 \\
 \texttt{+-0+} & s1RSB          & 0.0448          & 0.0491(1)           & -0.0044(1)         & 0.0445 \\
 \texttt{+-+0} & RS*            & 0.1304          & 0.0259(3)           & 0.1045(2)          &        \\
 \texttt{+100} & RS             & 0.2681          &                     &                    &        \\
 \texttt{+10+} & RS             & 0.2964          &                     &                    &        \\
 \texttt{+10-} & d1RSB          & 0.1038          & 0.0000              & 0.1038(2)          &        \\
 \texttt{+1+0} & RS             & 0.4584          &                     &                    &        \\
 \texttt{+1+-} & RS             & 0.2782          &                     &                    &        \\
 \texttt{+110} & RS             & 0.2617          &                     &                    &        \\
            \hline
\end{tabular}}
\caption{Classification of all non-equivalent outer-totalistic CSPs with homogeneous solutions for 3-regular graphs. The rules with only homogeneous solutions are indicated in Table~\ref{tbl:homogeneous rules}. RS* indicates that the hard-fields initialization yields a non-trivial solution while the small-noise initialization does not. The internal free entropy $\phi_{\rm int}$ and complexity $\Sigma$ and  are those obtained from the hard-field initialization at Parisi parameter $m=1$. $\phi_s$ indicates the free entropy for the rules presenting a static 1RSB phase.}
\label{tbl:classification_with_homogeneous}
\end{table}

\paragraph{Static 1RSB.}

The RS estimation of the free entropy density is wrong in the case of s1RSB. The rules \texttt{+-00} and \texttt{+-0+} are in the s1RSB phase. In that case, the static one-step replica symmetry free entropy density $\phi_s$ is found by computing the complexity as a function of the internal free entropy. This is done by varying the Parisi parameter between $0$ and $1$ (in our case we used 30 values of $m$), and fitting the obtained points with a function of the form $\Sigma(\phi_{\rm int})=a+b2^{\phi_{\rm int}}+c3^{\phi_{\rm int}}$ as done in \cite{zdeborova_phase_2007}. The intersection of this curve with $\Sigma=0$ gives the value of $\phi_s$. Figure~\ref{fig:complexity_curve} (right) shows the complexity versus internal free entropy curve for rule \texttt{+-00}. We also recall that the RS solution of rule \texttt{+-00} and \texttt{+-0+} is unstable (see the discussion on stability in Section \ref{sec:Replica symmetric solution using belief propagation}), indicating the presence of replica symmetry breaking. We recall that the annealed free entropy is equal to the RS free entropy in the case of regular graphs \cite{mora_geometrie_2007}. Thus, rules \texttt{+-00} and \texttt{+-0+} are the only rules in the $d=3$ case where according to our study the quenched free entropy is not equal to the annealed one.

The classification introduced in Section \ref{sec:Classification of outer-totalistic rules} does not change by considering the 1RSB solution. Thus, at least for $d=3$, the RS solution is sufficient to determine if solutions exist.

\begin{table}[ht]
\centering
\tiny
\setlength{\tabcolsep}{4pt}
\renewcommand{\arraystretch}{1.2}
\subfloat{
\begin{tabular}{|c|c|c|c|c|}
 \hline
            \multicolumn{5}{|c|}{Locally contradictory} \\
            \hline
            \hline
            rule & phase & $\phi_{\rm RS}$  & $\phi_{\rm int}, m=1$ & $\Sigma, m=1$ \\
            \hline
 \texttt{-000} & UNSAT & $-\infty$ &  &  \\
 \texttt{-00-} & UNSAT & $-\infty$ &  &  \\
 \texttt{-0-0} & UNSAT & $-\infty$ &  &  \\
 \texttt{-0--} & UNSAT & $-\infty$ &  &  \\
 \texttt{--00} & UNSAT & $-\infty$ &  &  \\
 \texttt{--0-} & UNSAT & $-\infty$ &  &  \\
 \texttt{---0} & UNSAT & $-\infty$ &  &  \\
 \texttt{----} & UNSAT & $-\infty$ &  &  \\
            \hline
            \multicolumn{5}{c}{} \\
            \hline
            \multicolumn{5}{|c|}{No solutions but not locally contradictory} \\
            \hline
            \hline
            rule & phase & $\phi_{\rm RS}$  & $\phi_{\rm int}, m=1$ & $\Sigma, m=1$ \\
            \hline
 \texttt{-010} & UNSAT & -0.1116          & 0.0000    & -0.1116(2)   \\
 \texttt{-01-} & UNSAT & -0.2027          & 0.0000    & -0.2027(2)   \\
 \texttt{-+-0} & UNSAT & -0.0774          & 0.0000    & -0.0774(3)   \\
 \texttt{-+--} & UNSAT & -0.1744          & 0.0000    & -0.1745(3)   \\
 \texttt{--+0} & UNSAT & -0.0303          & 0.0000    & -0.0308(2)   \\
 \texttt{--10} & UNSAT & -0.1733          & 0.0000    & -0.1733(2)   \\
 \texttt{-100} & UNSAT & -0.0012          & 0.0000    & -0.0012(2)   \\
 \texttt{-10-} & UNSAT & -0.2027          & 0.0000    & -0.2027(3)   \\
 \texttt{-1-0} & UNSAT & -0.2502          & 0.0000    & -0.2502(2)   \\
 \texttt{10-0} & UNSAT & -0.0760          & 0.0000    & -0.0760(2)   \\
 \texttt{1-00} & UNSAT & -0.0189          & 0.0000    & -0.0189(1)   \\
 \texttt{1--0} & UNSAT & -0.3466          & 0.0000    & -0.3466(2)   \\
            \hline
\end{tabular}}
\quad
\subfloat{
\begin{tabular}{|c|c|c|c|c|}
  \hline
  \multicolumn{5}{|c|}{Subexponentially many solutions} \\
            \hline
            \hline
            rule & phase & $\phi_{\rm RS}$ &  $\phi_{\rm int}, m=1$ & $\Sigma, m=1$  \\
            \hline
 \texttt{1010} & d1RSB & 0.0000 & 0.0000 & 0.0000(2)  \\
            \hline
            \multicolumn{5}{c}{} \\
            \hline
            \multicolumn{5}{|c|}{Exponentially many solutions} \\
            \hline
            \hline
            rule & phase & $\phi_{\rm RS}$ & $\phi_{\rm int}, m=1$ & $\Sigma, m=1$ \\
            \hline
 \texttt{-0+0} & RS* & 0.1016   & 0.0000                      & 0.1016(2)            \\
 \texttt{-0+-} & RS* & 0.0523   & 0.0000                      & 0.0523(2)            \\
 \texttt{-+00} & RS             & 0.2191 &                    &                      \\
 \texttt{-+0-} & RS             & 0.1541 &                    &                      \\
 \texttt{-++0} & RS             & 0.4880 &                    &                      \\
 \texttt{-++-} & RS             & 0.4055 &                    &                      \\
 \texttt{-+10} & RS* & 0.2149   & 0.0817(3)                   &  0.1331(3)           \\
 \texttt{-1+0} & RS             & 0.3551 &                    &                      \\
 \texttt{-110} & RS             & 0.1893 &                    &                      \\
 \texttt{1000} & RS             & 0.2617 &                    &                      \\
 \texttt{10+0} & RS* & 0.2845   & 0.1460(4)                   & 0.1384(4)            \\
 \texttt{1+00} & RS             & 0.4414 &                    &                      \\
 \texttt{1++0} & RS             & 0.5829 &                    &                      \\
 \texttt{1+-0} & d1RSB          & 0.1038 & 0.0000             & 0.1038(2)            \\
 \texttt{1100} & RS             & 0.2350 &                    &                      \\
            \hline
\end{tabular}}
\caption{Classification of all non-equivalent outer-totalistic CSPs without homogeneous solutions for 3-regular graphs. RS* indicates that the hard-fields initialization yields a non-trivial solution while the small-noise initialization does not. The internal free entropy $\phi_{\rm int}$ and complexity $\Sigma$ and are those obtained from the hard-field initialization at Parisi parameter $m=1$. In the case of rules with no homogeneous solutions for $d=3$, no rule presents a static 1RSB phase.}
\label{tbl:classification_without_homogeneous}
\end{table}

\paragraph{Stability of the 1RSB solution}
Checking the validity of the 1RSB solution is more delicate than in the RS case. We only numerically check the so-called \textit{type II} stability \cite{montanari_instability_2004, zdeborova_statistical_2008}, which is done by adding noise to the obtained messages of the population dynamics. We add a noise $|\eta|, \eta\sim \mathcal{N}(0, 10^{-8})$ and normalize the messages. If the population dynamics starting from the noisy messages converges back to the previous non-noisy messages, we say that the solution is stable. If the noisy messages go to the RS fixed point, we are in the case where the small noise reconstruction is not possible, and we are in the `fake cluster' case discussed above. However, if the noisy messages converge to another non-trivial fixed point, there is an instability of type II which indicates that further steps of replica symmetry breaking are required. This last case did not arise, giving an indication that the 1RSB result is correct.

\section{Freezing and Computational Hardness}\label{sec:Freezing and computational hardness}

It has been proven that a large class of algorithm fail to find solutions to problems presenting the overlap gap property (OGP) \cite{gamarnik_limits_2017, gamarnik_overlap_2021}. Informally, a problem presents OGP 
if the clusters of solutions are well separated, in the sense that there are no pairs of solutions at a certain intermediate distance.
Works in statistical physics put forward a conjecture that solutions within frozen clusters (a weaker condition than OGP) are algorithmically difficult to find (e.g. require exponential time) \cite{zdeborova_constraint_2008, huang_origin_2014, zdeborova_statistical_2016, gamarnik_disordered_2022}. It is also believed that solutions to problems that are replica symmetric can be found in polynomial time. We provide experimental evidence for these conjectures by solving outer-totalistic CSPs using the belief propagation reinforcement algorithm. The algorithm is presented below. 

It is important to keep in mind that we only characterized the thermodynamically relevant clusters, which contain \textit{typical solutions}. However, it is possible that non-dominating clusters have very different properties in terms of the hardness of finding solutions that belong to them.  
It has been shown that in the presence of atypical clusters with large local entropy, solutions can still be found in polynomial time \cite{dallasta_entropy_2008, abbe_binary_2021}. One such case arises for a rule of degree $d=4$ and is discussed below.

\paragraph{Belief propagation reinforcement} The belief propagation reinforcement algorithm, introduced in \cite{chavas_survey-propagation_2005}, is a solver based on BP. Given a graph and a rule, the solver tries to produce a solution. The main idea is to reinforce, i.e. to bias, the variables in the direction of the marginals obtained by BP. The BP-reinforcement equation reads
\begin{equation}\label{eq:BP_reinforcement}
    \psi^{i\rightarrow j}_{s_i, s_j}=
    \frac{1}{Z^{i\rightarrow j}}\sum_{\{s_k\}_{k\in \partial i \setminus j}} f(s_i, s_j, \{s_k\}_{k\in\partial i\setminus j}) \prod_{k\in \partial i \setminus j} b^{(k)}_{s_k} \psi_{s_k, s_i}^{k\rightarrow i}.
\end{equation}
where $b^{(k)}_{s_k}$ is the bias. To update the bias, we use the same heuristics as in \cite{zdeborova_constraint_2008}:
\begin{equation}\label{eq:update_bias}
\begin{split}
    b_{1}^{(i)}=\pi,\,   b_{0}^{(i)}=1-\pi,\, \text{if} \, \chi_{0}^{(i)}>\chi_{1}^{(i)} \\
    b_{1}^{(i)}=1-\pi,\,   b_{0}^{(i)}=\pi,\, \text{if} \, \chi_{0}^{(i)}\leq\chi_{1}^{(i)}
\end{split}
\end{equation}
where $0\leq\pi\leq 1/2$ is a hyperparameter and $\chi_{s_i}^{(i)}$ is the BP estimate of the marginal probability that node $i$ is in state $s_i$. $\chi_{s_i}^{(i)}$ is given by
\begin{equation}
    \chi_{s_i}^{(i)}=\frac{1}{Z^{\text{site}}_\chi}\prod_{k\in\partial i}\left(\chi_{0, s_i}^{k\rightarrow i}+\chi_{1, s_i}^{k\rightarrow i}\right)
\end{equation}
where $Z^{\text{site}}_\chi$ is the normalization. $\chi_{s_i, s_j}^{i\rightarrow j}$ are the BP estimate of the marginal probability that the variable node $ij$ in the factor graph takes values $(s_i, s_j)$. This is given by
\begin{equation}
    \chi_{s_i, s_j}^{i\rightarrow j}=\frac{\psi_{s_i, s_j}^{i\rightarrow j}\psi_{s_j, s_i}^{j\rightarrow i}}{Z_{\chi}}
\end{equation}
where $Z_{\chi}$ is the normalization. Each bias $b_{s_i}^{(i)}$ is updated following \eqref{eq:update_bias}, but the update is done only with probability 
\begin{equation}
    p(t)=1-(1+t)^{-\gamma}
\end{equation}
where $t$ is the iteration step and $\gamma$ a hyperparameter. We fix $\gamma=0.1$ in our case. To obtain a possible solution $\mathbf{s}$ from the algorithm, we take
\begin{equation}
    s_i=\operatorname{argmax}_{s_l} b_{s_l}^{(i)}\,\,\forall i=1,..., N.
\end{equation} The pseudocode is given in Algorithm \ref{algo:BP_reinforcement}. In practice, $\pi$ is found by doing a grid search starting with $\pi=0.5$ and decreasing it until a solution is found or $\pi=0$ is reached. In that last case, the algorithm failed to find a solution. We use $25$ values of $\pi$ equidistant between $0$ and $0.5$, and a maximum number of iterations $T=10^4$. Each time the algorithm is restarted, the bias and the messages are initialized at random.

\begin{algorithm}[h!]
\caption{BP-REINFORCEMENT $(G, T, \gamma, \pi, \epsilon)$}
\label{algo:BP_reinforcement}
\begin{algorithmic}[1]
    \STATE Initialize the bias $b_{s_i}^i$ and the messages $\psi_{s_i, s_j}^{i\rightarrow j}$ randomly on the factor graph of $G$ and normalize them.
    \STATE $t \leftarrow 0$
    \STATE Compute the current configuration $s_i=\operatorname{argmax}_{s_l} b_{s_l}^i$
    \WHILE{$\mathbf{s}$ is not a solution and $t \leq T$}
        \STATE Update all the messages $\psi_{s_i, s_j}^{i\rightarrow j}$ according to \eqref{eq:BP_reinforcement} with dampening $\epsilon$.
        \STATE Update every bias $b_{s_i}^{(i)}$ with probability $p(t)$ according to \eqref{eq:update_bias}
        \STATE Update $s_i=\operatorname{argmax}_{s_l} b_{s_l}^{(i)}$
        \STATE $t \leftarrow t+1$
    \ENDWHILE
\end{algorithmic}
\end{algorithm}

\paragraph{d=3 rules} We focus on the problems that have $\phi\geq 0$ with no homogeneous solutions, since the rules that have a homogeneous solution are trivial to solve. In the case $d=3$, we tried to find solutions on $10$ different graphs of size $N=10^5$ for each rule. Solutions were found for all $10$ graphs for all rules except for the 1RSB rules with frozen variables \texttt{1+-0}, where a solution on only one graph was found (which we attribute to a finite size effect). For the rule \texttt{1010} no solutions were found. This is shown in Table \ref{tbl:hardness} (left). These results are in agreement with the conjecture that frozen solutions are hard to find for robust algorithms, and are also coherent with the hypothesis that the `fake clustered' rules (marked RS*) are algorithmically easy. We also found no solution for the rules with negative free entropy density but that are not locally contradictory. 

Note the special case of rule \texttt{1010}, which can be mapped to a system of linear equations modulo $2$ and can thus be solved in polynomial time. Using Gaussian elimination, we found a solution for each of the $10$ graphs of size $8\cdot10^4$ that we tested. Rules \texttt{+---}, \texttt{-+--}, \texttt{+-+-}, \texttt{--10}, \texttt{-10-}, \texttt{0101} and \texttt{0-10} can also be mapped to a system of linear equations over finite fields, but are of lesser interest presenting either a trivial homogeneous solution or no solutions at all. The \texttt{1010} rule is a parity check and is known to be solvable in polynomial time. A similar situation arises in XORSAT problems, where the space of solutions also exhibits clusters and frozen variables \cite{cocco_rigorous_2003, mora_geometrical_2006} while the problem is solvable in polynomial time \cite{barthel_hiding_2002, haanpaa_hard_2006}. However, note that the algebraic structure allowing Gaussian elimination disappears if one adds noise to the problem \cite{barthel_hiding_2002}.
If we focus on algorithms robust to noise the problem \texttt{0101} will also be algorithmically hard.

\begin{table}
\centering
\scriptsize
\setlength{\tabcolsep}{-16pt}
\addtolength{\tabcolsep}{0em}
\renewcommand{\arraystretch}{1.2}
\subfloat{
\begin{tabular}{ @{\hspace{2pt}} lcc @{\hspace{2pt}} }
& \multicolumn{1}{l}{ \hspace{14pt} Satisfiable $d=3$ rules with no homogeneous solutions} 
\\ \toprule
Rule & Phase & Rate of success\\
    \midrule
 \texttt{-0+0} & RS* & 1\\
 \texttt{-0+-} & RS*& 1\\
 \texttt{-+00} & RS & 1\\
 \texttt{-+0-} & RS & 1\\
 \texttt{-++0} & RS & 1\\
 \texttt{-++-} & RS & 1\\
 \texttt{-+10} & RS*& 1\\
 \texttt{-1+0} & RS & 1\\
 \texttt{-110} & RS & 1\\
 \texttt{1000} & RS & 1\\
 \texttt{10+0} & RS* & 1\\
 \texttt{1+00} & RS & 1\\
 \texttt{1++0} & RS & 1\\
 \texttt{1+-0} & d1RSB & 0.1\\
 \texttt{1100} & RS & 1\\
 \texttt{1010} & d1RSB & 0\\
 \bottomrule
\end{tabular}
}
\qquad\qquad
\subfloat{
\setlength{\tabcolsep}{0pt}
\begin{tabular}{lc}
& \multicolumn{1}{c}{\shortstack[l]{\hspace{-18pt}Rigid d1RSB $d=4$ rules\\ \hspace{-18pt}with no homogeneous solution}}
\\ \toprule
Rule &  Rate of success\\
    \midrule
 \texttt{-0+-0}  & 0.5\\
 \texttt{-0+--} &  0\\
 \texttt{-+-+0}  & 0\\
 \texttt{-+-+-}  & 0\\
 \texttt{-+10-} & 0\\
 \texttt{--+00} & 0\\
 \texttt{-1000} & 0\\
 \texttt{-10+0} & 0\\
 \texttt{-1+-0} & 0\\
 \texttt{-1-+0} & 0\\
 \texttt{10100} & 0\\
 \texttt{1+-00} & 1\\
 \bottomrule
\end{tabular}
}
\caption{Left: Rate of success of the belief propagation reinforcement algorithm on random 3-regular graphs with $N=10^5$ nodes for 10 different graphs. The studied rules have $\phi\geq 0$ and no homogeneous solutions. Right: Rate of success using the belief propagation reinforcement algorithm on random 4-regular graphs with $N=10^5$ nodes for 10 different graphs for the stable and frozen d1RSB rules with no homogeneous solutions.}
\label{tbl:hardness}
\end{table}

\paragraph{d=4 rules and local entropy} We investigate the case $d=4$, for which we identify $12$ stable d1RSB rules with no homogeneous solutions and containing frozen variables out of the possible 528 non-equivalent rules. Using again BP reinforcement on $10$ graphs of size $N=10^5$, we find no solution except for rule \texttt{-0+-0} for which we found solutions on 5 out of the 10 graphs, and rule \texttt{1+-00} for which a solution was found on every tested graph, see Table \ref{tbl:hardness} (right). 

As discussed before, we expect the found solutions to belong to non-dominating clusters with large local entropy. To investigate this, \cite {baldassi_subdominant_2015} or more recently \cite{baldassi_typical_2023} study the \textit{local entropy density} $s_{LE}(\mathbf{s}, d)$, which is the logarithm of the number of solutions at a given (normalized) Hamming distance $d$ from a given solution $\mathbf{s'}$ (note the difference with the entropy of the \textit{Franz-Parisi potential} \cite{franz_recipes_1995}, which is equivalent to the local entropy in the case where the given solution $\mathbf{s}'$ is typical). They argue that solutions in wide and flat connected regions of solutions can be algorithmically easy to find. These regions are characterized by the absence of a gap in their local entropy. More precisely, we say that there is a gap if there exist two distances $\alpha>0$ and $\beta>0$ with $\alpha<\beta$ such that $\phi_{LE}(\mathbf{s}, \alpha)\geq 0,\, \phi_{LE}(\mathbf{s}, \beta)\geq 0$ and for some $\gamma\in[\alpha, \beta]$ we have $\phi_{LE}(\mathbf{s}, \gamma)<0$. We compute the local entropy within the RS approximation using belief propagation on a given graph by adding a Lagrange multiplier $\kappa$ tuning the distance to a given solution. The studied partition function then is
\begin{equation}\label{eq:Partition function local entropy}
    Z_{LE}(\mathbf{s'}, d)=\sum_{\mathbf{s}\in\mathcal{C}}\prod_{i=1}^N e^{\kappa \mathds{1}(s_i\neq s'_i)} f(s_i, \{s_j\}_{j\in\partial i}) f(s'_i, \{s'_j\}_{j\in\partial i})
\end{equation}
where $\mathds{1}$ is the indicator function that is $1$ if the argument is true and is $0$ otherwise. The local free entropy density is $\phi_{LE}(\mathbf{s'}, d)=\log(Z_{LE}(\mathbf{s'}, d))/N$. The local entropy is obtained by a Legendre transform and is given by $s_{LE}=\phi_{LE}-\kappa d$. The BP equation, the distance and the local entropy density are obtained following a similar procedure as in Appendix~\ref{appendix:derivation BP on graphs}. To obtain the local entropy density at a chosen distance, the Lagrange multiplier $\kappa$ is tuned dynamically during the BP iterations. The derivation of the BP equation and the iteration algorithm is presented in more details in Appendix~\ref{appendix:local entropy}. 

Figure~\ref{fig:LE} (left) presents the local entropy density as a function of the distance to a solution found by using BP reinforcement on a graph of size $N=10^5$ for the rule \texttt{1+-00}. We notice that there is no gap, indicating a flat and wide minima. This also indicates that the solution is atypical, since typical solutions are clustered and thus present a gap. However, recall that the RS free entropy is an upper bound of the true free entropy, which is not necessarily reached, so a gap might emerge by considering further steps of replica symmetry breaking. 

On the other hand, Figure~\ref{fig:LE} (right) shows the local entropy density for rule \texttt{1+-0}. Since finding solutions for this rule on large graphs is hard, we use $N=1000$ where solutions could be consistently found. We notice the presence of a gap, as expected for a solution within a cluster. However, note that the gap is small and the obtained curve can slightly change depending on the generated graph, probably due to its small number of nodes. We also checked the unique solution found for $N=10^5$, and it does not present a gap in its local entropy. We suppose that this solution was found from an atypical generation of the graph due to finite $N$. The solutions of rule \texttt{-0+-0} also do not present a gap in their local entropy. However, solutions are found more frequently in this case which means that the explanation of atypical graphs is unlikely. It is possible that, in this case, sub-dominant dense clusters (in which solutions have a large local entropy) exist, but that the solutions are not always found by our choice of hyperparameters in the BP reinforcement algorithm.

\begin{figure}[t]
  \centering
  \begin{subfigure}{0.48\textwidth}
    \centering
    \includegraphics[width=\textwidth]{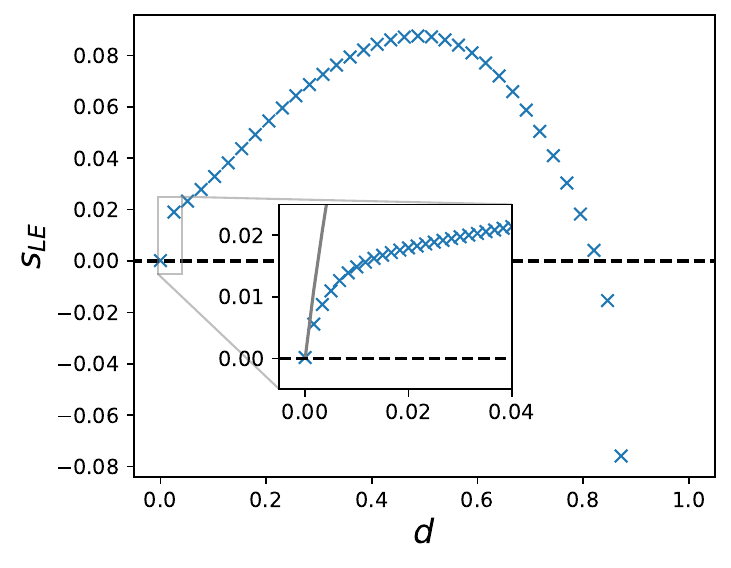} 
  \end{subfigure}
  \hfill
  \begin{subfigure}{0.48\textwidth}
    \centering
    \includegraphics[width=\textwidth]{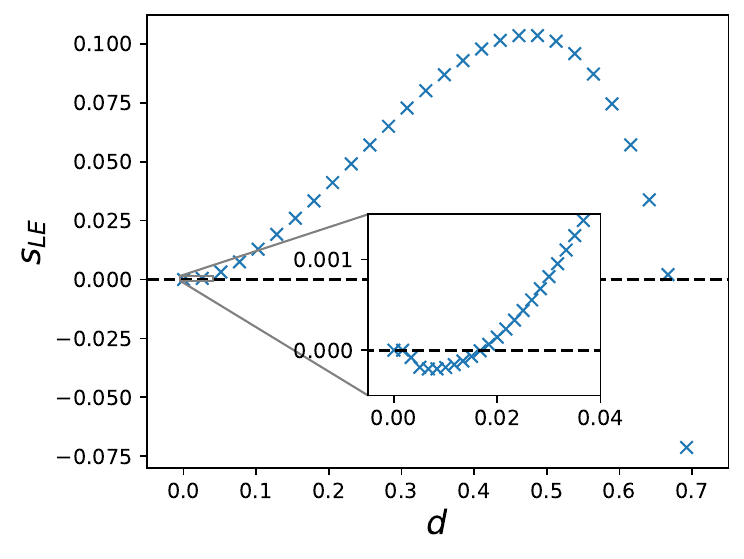} 
  \end{subfigure}
  \caption{Left: Local entropy $S_{LE}$ as function of the normalized distance $d$ for an empirical solution for rule \texttt{1+-00}, $N=10^5$. The grey line in the inset is the entropy with no constraints. While typical solutions are frozen, the found solution is in a subdominant cluster with large local entropy (for small $d$ the local entropy is close to the entropy with no constraints) and contains no frozen variables. Right: Local entropy as a function of the normalized distance for an empirical solution containing frozen variables for rule \texttt{1+-0}, $N=1000$. Notice the presence of a gap in the zoomed-in region close to $0$.}
  \label{fig:LE}
\end{figure}

\section*{Conclusion}
In this work, we provided the cavity equation under both the replica symmetric and 1-step replica symmetry-breaking assumptions for all outer-totalistic constraint satisfaction problems on $d$-regular graphs. By numerically solving these equations, we determined the existence and number of solutions. The 1RSB calculation also allowed to characterize the structure of the space of typical solutions. Additionally, the computational hardness of the problems was explored experimentally using the belief propagation reinforcement algorithm.  This analysis uncovered the following interesting phenomena:
\begin{itemize}
\item For the $d=3$ rules, we find that the annealed free entropy is equal to the quenched free entropy with the exception of rules \texttt{+-00} and \texttt{+-0+} which present a static 1-step replica symmetry breaking. Thus, for these two rules, the quenched free entropy is strictly lower than the annealed one.
\item For the $d=3$ rules, we uncovered one rule, \texttt{1+-0}, where solutions are hard to find algorithmically. This rule presents clustering and frozen variables.
\item In the $d=4$ case, multiple rules were found to present clustering and frozen variables and no trivial homogeneous solution. While no solutions were found for most of them, solutions to rule \texttt{1+-00} were easy to find. An analysis of the local entropy near the found solutions indicates that these solutions belong to a subdominant dense cluster, which does not contain frozen variables.
\end{itemize}

Several extensions of this work could be interesting to consider. While we limited our study to a binary alphabet, expanding our approach to more states is straightforward using our formalism, which would then also include problems such as graph-coloring. Additionally, investigating rules that go beyond the nearest neighbors could also yield interesting insights. Concerning the typical algorithmic complexity, the overlap gap property could be studied using the cavity method. Indeed, the overlaps can be estimated using this method, which could potentially provide additional insights into the algorithmic hardness of certain problems. Finally, the outer-totalistic CSPs were introduced in this work through the lens of stationary solutions of cellular automata on graphs. Going beyond stationarity to study the dynamics of graph cellular automata is feasible with the cavity method as shown in recent works \cite{behrens_backtracking_2023, behrens_dynamical_2023}. Extending this line of work could prove an exciting avenue into the dynamic properties of complex systems.

\section*{Acknowledgements}
We thank Vittorio Erba and Barbora Hudcov\'a for insightful and helpful discussions.

\bibliography{bibliography}

\newpage

\appendix

\section{Non-equivalent Outer-totalistic Rules}\label{appendix:Non-equivalent outer-totalistic rules}

We want to remove the ``empty $\leftrightarrow$ occupied" symmetry that produces equivalent rules. Extending the procedure of \cite{marr_outer-totalistic_2009}, the operator $\mathcal{T}: \xi \rightarrow 1-\xi, \, \xi\in\{0,1\}$, which switches 0s and 1s, is introduced. Abusing the notation, $\mathcal{T}$ can be applied to a configuration $\mathbf{s}$ using the notation $\mathcal{T}\mathbf{s}$, e.g. $\mathcal{T}(0,1,0,1,1)=(1,0,1,0,0)$. $\mathcal{T}$ can also be applied to a rule $\alpha_0 \alpha_1 \hdots \alpha_d$ with $\alpha_0, \alpha_1, \hdots, \alpha_d \in \{\texttt{0,1,+,-}\}$ using the notation $\mathcal{T}(\alpha_0 \alpha_1 \hdots \alpha_d)$, where the elements $\{\texttt{+,-}\}$ are invariant under $\mathcal{T}$. For example, $\mathcal{T}(\texttt{0+1-})=\texttt{1+0-}$. We consider here that the rule is describing the dynamics for the next time step: \texttt{+} indicates that the node stays in the same state, \texttt{-} that it switches state, \texttt{0} that is becomes $0$ and \texttt{1} that it becomes 1. One time-step update of a configuration $\mathbf{s}$ under rule $\alpha_0 \alpha_1 \hdots \alpha_d$ is denoted $\alpha_0 \alpha_1 \hdots \alpha_d \cdot\mathbf{s}$. It can be observed that $\alpha_0 \alpha_1 \hdots \alpha_d\cdot\mathbf{s}=\mathcal{T}\left(\mathcal{T}(\alpha_d \hdots \alpha_1 \alpha_0)\cdot \mathcal{T}\mathbf{s}\right)$ where $\alpha_d \hdots \alpha_1 \alpha_0$ is $\alpha_0 \alpha_1 \hdots \alpha_d$ in reverse order. Thus, it can be said that the rule $\mathcal{T}(\alpha_d \hdots \alpha_1 \alpha_0)$ is symmetric to the rule $\alpha_0 \alpha_1 \hdots \alpha_d$. Notice that some rules are self-symmetric, e.g. rule \texttt{0-1}.

If each node has two neighbors, a total of $4^3=64$ outer-totalistic rules exist. 8 of them are self-symmetric: \texttt{0+1}, \texttt{1+0}, \texttt{0-1}, \texttt{1-0}, \texttt{+++}, \texttt{+-+}, \texttt{-+-} and \texttt{---}. All the other rules have exactly one different symmetric rule. Thus, a total of $8+(64-8)/2=36$ nonequivalent outer-totalistic rules are obtained\footnote{Note that \cite{marr_outer-totalistic_2009} only identifies 34 nonequivalent outer-totalistic rules using the same method. The authors were contacted about this discrepancy, and came to the conclusion that it was a typo.}. This procedure can easily be extended to identify the nonequivalent rules for $d>2$.

\section{Derivation of the BP Equation and Observables on a Tree}\label{appendix:derivation BP on graphs}
In this appendix, we give a detailed derivation of eq. \eqref{eq:BP} and \eqref{eq:RS_free_entropy} starting from the graphical model presented in Section \ref{sec:Replica symmetric solution using belief propagation}. We will treat a more general case than the probability distribution defined in eq. \eqref{eq:probability_measure_stationary_configuration}:
\begin{equation}\label{eq:probability_distribution_mu}
    P_G(\mathbf{s})
    =
    \frac{\prod_{i=1}^N e^{\mu s_i}f(s_i, \{s_j\}_{j\in\partial i})}{Z_G}.
\end{equation}
where $\mu$ is the \textit{chemical potential}, a Lagrange multiplier allowing us to tune the density of occupied nodes $r=\sum_{i=1}^N s_i/N$. Note that additional Lagrange multipliers can easily be added to compute other quantities of interest, such as the local entropy as done in Section \ref{sec:Freezing and computational hardness}. To obtain eq.\eqref{eq:BP} and \eqref{eq:RS_free_entropy}, simply set $\mu=0$.

\begin{figure}
    \centering
    \includegraphics[scale = 0.4]{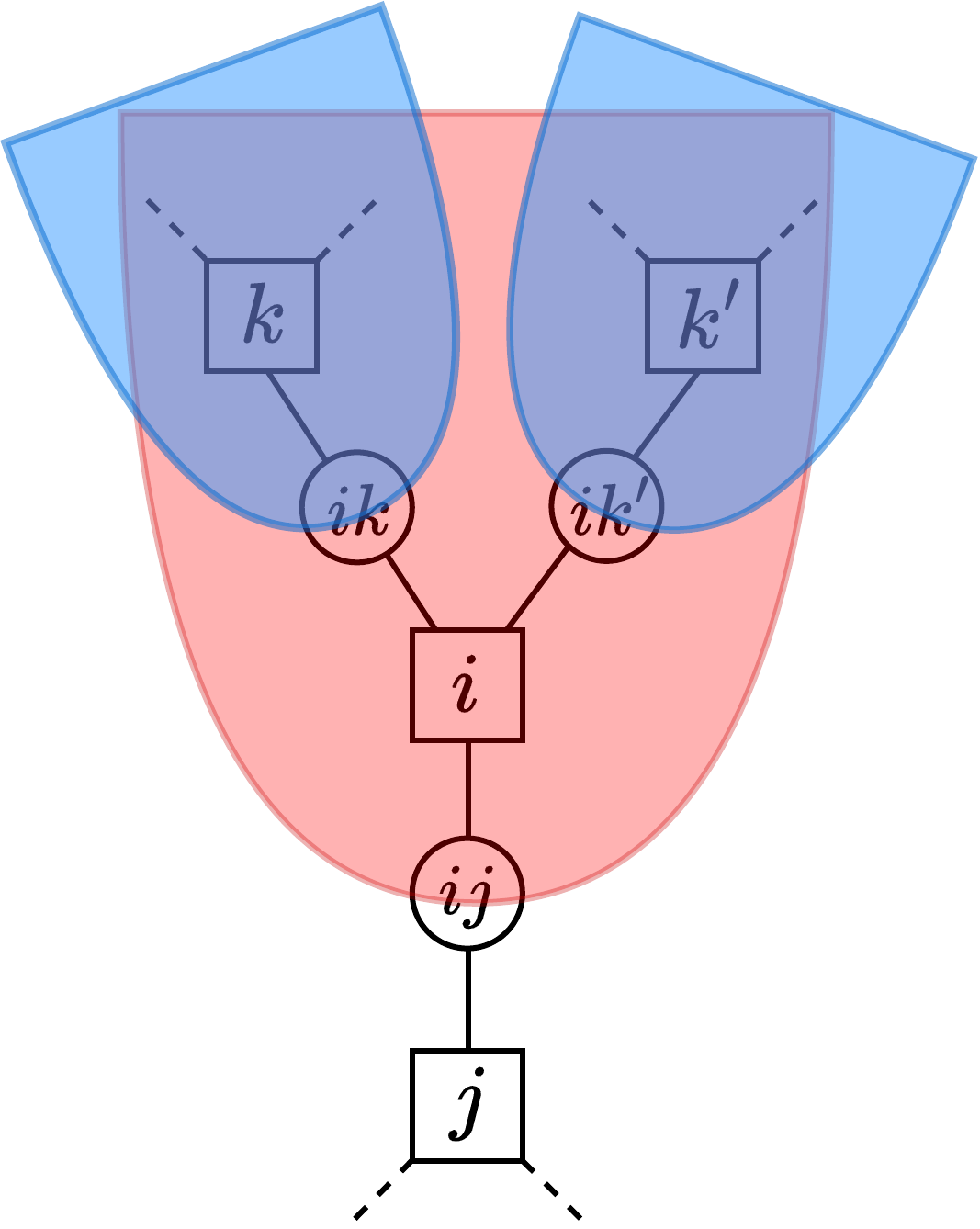}
    \caption{Factor graph with the subgraph $G^{i\rightarrow j}$ in red and the subgraphs $\{G^{k\rightarrow i}\}_{k\in\partial i\setminus j}$ in blue.}
    \label{fig:auxiliary_partition_function}
\end{figure}

\paragraph{BP equation.} Suppose that the original graph is a tree. From our construction of the graphical model, this implies that the factor graph is also a tree. We define the \textit{auxiliary partition function} $V_{s_i,s_j}^{i\rightarrow j}$ as being the partition function of the system above $ij$ with the values of $s_i, s_j$ fixed. This corresponds to all the variables in the branch of the tree that include $i$ but with the edge $ij$ removed. Let us name this sub-graph $G^{i\rightarrow j}$. $G^{i\rightarrow j}$ is shown in red in Figure~\ref{fig:auxiliary_partition_function}. The auxiliary partition function reads
\begin{equation}
    V_{s_i,s_j}^{i\rightarrow j}=\sum_{\{s_l\}_{l\in G^{i\rightarrow j}}}\prod_{l\in G^{i\rightarrow j}}e^{\mu s_l} f(s_l, \{s_m\}_{m\in\partial l}).
\end{equation}

Note that we indicate the neighborhood on the original graph for simplicity. Supposing that the factor graph is a tree, fixing the value of variable $s_i, s_j$ makes the branches above $i$ independent of each other. Let us name these branches $G^{k\rightarrow i}$ with $k\in\partial i\setminus j$. They are shown in blue in Figure~\ref{fig:auxiliary_partition_function}. Thus, we can split the sum according to the branch they belong to into a recursive relation:
\begin{equation}\label{eq:recursion_relation}
\begin{split}
     V_{s_{i,j}}^{i\rightarrow j}&=\sum_{\{s_k\}_{k\in \partial i \setminus j}} e^{\mu s_i} f(s_i, s_j, \{s_k\}_{k\in\partial i\setminus j})
     \\&
     \,\,\,\,\,\,\,\,\,\,\,\,\times \prod_{k\in \partial i \setminus j} \sum_{\{s_l\}_{l\in G^{k\rightarrow i}}}\prod_{l \in G^{k\rightarrow i}}e^{\mu s_l} f(s_l, \{s_m\}_{m\in\partial l})
     \\
     &=\sum_{\{s_k\}_{k\in \partial i \setminus j}} e^{\mu s_i} f(s_i, s_j, \{s_k\}_{k\in\partial i\setminus j}) \prod_{k\in \partial i \setminus j}  V_{s_{k,i}}^{k\rightarrow i}.
 \end{split}
\end{equation}
Using \eqref{eq:recursion_relation} starting from the leaves, one can obtain all the auxiliary partition functions of the tree.

Define the \textit{message} $\psi^{i\rightarrow j}_{s_{i,j}}$ as the normalized auxiliary partition function
\begin{equation}\label{eq:def_messages}
    \psi^{i\rightarrow j}_{s_i, s_j}:=\frac{ V_{s_i,s_j}^{i\rightarrow j}}{\sum_{s_i,s_j}  V_{s_i,s_j}^{i\rightarrow j}},
\end{equation}
where $\sum_{s_i,s_j}$ is the sum over all possible values of $s_i$ and $s_j$, e.g. in our case $s_i=s_j=0$, $s_i=0$ and $s_j=1$, $s_i=1$ and $s_j=0$, $s_i=s_j=1$.
Messages are more convenient to work with than the auxiliary partition function, since they do not scale exponentially in $N$ and represent the probability that variable $ij$ takes value $s_i,s_j$ when restricted to the sub-graph $G^{i\rightarrow j}$ (the red part of Figure~\ref{fig:auxiliary_partition_function}). Using the recursion relation \eqref{eq:recursion_relation} and the definition \eqref{eq:def_messages}, we can also obtain a recursion relation for the messages:
\begin{equation}
\begin{split}
    \psi^{i\rightarrow j}_{s_i,s_j}&=
    \frac{\sum_{\{s_k\}_{k\in \partial i \setminus j}} e^{\mu s_i} f(s_i, s_j, \{s_k\}_{k\in\partial i\setminus j})\prod_{k\in \partial i \setminus j}  V_{s_k, s_i}^{k\rightarrow i}}{\sum_{s_i, s_j} \sum_{\{s_k\}_{k\in \partial i \setminus j}} e^{\mu s_i} f(s_i, s_j, \{s_k\}_{k\in\partial i\setminus j}) \prod_{k\in \partial i \setminus j}  V_{s_k, s_i}^{k\rightarrow i}}
    \\
    & \,\,\,\,\,\,\,\, \times \overbrace{\frac{\prod_{k\in \partial i \setminus j} \sum_{s_k, s_i} V_{s_k, s_i}^{k\rightarrow i}}{\prod_{k\in \partial i \setminus j} \sum_{s_k, s_i} V_{s_k,s_i}^{k\rightarrow i}}}^{=1}
    \\&= \frac{\sum_{\{s_k\}_{k\in \partial i \setminus j}} e^{\mu s_i} f(s_i, s_j, \{s_k\}_{k\in\partial i\setminus j}) \prod_{k\in \partial i \setminus j}  \psi_{s_k, s_i}^{k\rightarrow i}}{\sum_{s_i,s_j} \sum_{\{s_k\}_{k\in \partial i \setminus j}} e^{\mu s_i} f(s_i, s_j, \{s_k\}_{k\in\partial i\setminus j}) \prod_{k\in \partial i \setminus j}  \psi_{s_k,s_i}^{k\rightarrow i}}
\end{split}
\end{equation}
Noticing that
\begin{equation}
    Z^{i\rightarrow j}:=\sum_{s_i,s_j} \sum_{\{s_k\}_{k\in \partial i \setminus j}} e^{\mu s_i} f(s_i, s_j, \{s_k\}_{k\in\partial i\setminus j}) \prod_{k\in \partial i \setminus j}  \psi_{s_{(ki)}}^{k\rightarrow i}
\end{equation}
is a normalization, we obtain the following self-consistent equation for the message:
\begin{equation}\label{eq:BP_equations}
    \psi^{i\rightarrow j}_{s_i, s_j}=
    \frac{1}{Z^{i\rightarrow j}}\sum_{\{s_k\}_{k\in \partial i \setminus j}} e^{\mu s_i} f(s_i, s_j, \{s_k\}_{k\in\partial i\setminus j}) \prod_{k\in \partial i \setminus j} \psi_{s_k, s_i}^{k\rightarrow i}.
\end{equation}
This is the \textit{Belief Propagation equation} given in eq. \eqref{eq:BP_equations_graph}. As for the auxiliary partition functions, one can obtain recursively each message by starting from the leaves of the tree.

\paragraph{Observables.} 
Starting from the leaves, we can collect recursively the contribution of each branch to the partition function of the original problem to obtain the full partition function:
\begin{equation}\label{eq:Z_auxiliary}
    Z_G=\sum_{s_r}\sum_{\{s_k\}_{k\in\partial r}}e^{\mu s_r} f(s_r, \{s_k\}_{k\in\partial r}) \prod_{k\in\partial r} V^{k\rightarrow r}_{s_k, s_r}
\end{equation}
where $r$ indicates a root of the tree. We can write this expression as a function of the messages instead of the auxiliary partition function:
\begin{equation}
\begin{split}
    Z_G&=Z_G\overbrace{\frac{\prod_{i=1, i\neq r}^N\sum_{s_i}\sum_{s_j}V_{s_j, s_i}^{j\rightarrow i}V_{s_i,s_j}^{i\rightarrow j}}{\prod_{(ij)\in E(G)}\sum_{s_i}\sum_{s_j}V_{s_j,s_i}^{j\rightarrow i}V_{s_i,s_j}^{i\rightarrow j}}}^{= 1}
    \\&=Z_G\frac{\prod_{i=1, i\neq r}^N\sum_{s_i}\sum_{\{s_k\}_{k\in\partial i}} e^{\mu s_i} f(s_i, \{s_k\}_{k\in\partial i})\prod_{k\in\partial i}V_{s_k, s_i}^{k\rightarrow i}}{\prod_{ij\in E(G)}\sum_{s_i}\sum_{s_j}V_{s_j, s_i}^{j\rightarrow i}V_{s_i, s_j}^{i\rightarrow j}}
    \\&=\frac{\prod_{i=1}^N\sum_{s_i}\sum_{\{s_k\}_{k\in\partial i}} e^{\mu s_i} f(s_i, \{s_k\}_{k\in\partial i})\prod_{k\in\partial i}V_{s_k, s_i}^{k\rightarrow i}}{\prod_{ij\in E(G)}\sum_{s_i}\sum_{s_j}V_{s_j, s_i}^{j\rightarrow i}V_{s_i, s_j}^{i\rightarrow j}}
    \\&=\left(\prod_{i=1}^N\frac{\sum_{s_i}\sum_{\{s_k\}_{k\in\partial i}} e^{\mu s_i} f(s_i, \{s_k\}_{k\in\partial i}) \prod_{k\in\partial i}V_{s_k, s_i}^{k\rightarrow i}}{\prod_{k\in \partial i}\sum_{s_k, s_i} V_{s_k,s_i}^{k\rightarrow i}}\right)
    \\&\,\,\,\,\,\,\,\,\,\, \times \left(\prod_{ij\in E(G)}\frac{\sum_{s_i}\sum_{s_j}V_{s_j, s_i}^{j\rightarrow i}V_{s_i, s_j}^{i\rightarrow j}}{\sum_{s_i, s_j} V_{s_i, s_j}^{i\rightarrow j} \sum_{s_j, s_i}V_{s_j, s_i}^{j\rightarrow i}}\right)^{-1}
    \\&=\frac{\prod_{i=1}^N\sum_{s_i}\sum_{\{s_k\}_{k\in\partial i}}e^{\mu s_i} f(s_i, \{s_k\}_{k\in\partial i})\prod_{k\in\partial i}\psi_{s_k, s_i}^{k\rightarrow i}}{\prod_{ij\in E(G)} \sum_{s_i}\sum_{s_j}\psi_{s_i, s_j}^{i\rightarrow j} \psi_{s_j, s_i}^{j\rightarrow i}}
\end{split}
\end{equation}
where $E(G)$ denotes the set of edges of $G$. The property of the first line can be seen if one considers a tree with $i\rightarrow j$ going in the direction of the root. To go from the 1st to the 2nd line, we used the recursion equation \eqref{eq:recursion_relation}. From the 2nd to 3rd line, we see from \eqref{eq:Z_auxiliary} that $Z_G$ can be absorbed in the product over the nodes. We multiplied line 3 by $(\prod_{k\in \partial i}\sum_{s_k, s_i} V_{s_k, s_i}^{k\rightarrow i})(\sum_{s_i,s_j} V_{s_i,s_j}^{i\rightarrow j} \sum_{s_j,s_i}V_{s_j,s_i}^{j\rightarrow i})^{-1}$ (which is equal to $1$) to get to line 4. In the last equality we simply used the definition of the message \eqref{eq:def_messages}.

Defining
\begin{equation}\label{eq:Z^i}
    Z^i=\sum_{s_i}\sum_{\{s_k\}_{k\in\partial i}}e^{\mu s_i} f(s_i, \{s_k\}_{k\in\partial i})\prod_{k\in\partial i}\psi_{s_k,s_i}^{k\rightarrow i},
\end{equation}
\begin{equation}\label{eq:Z^ij}
    Z^{ij}=\sum_{s_i}\sum_{s_j}\psi_{s_i, s_j}^{i\rightarrow j} \psi_{s_j,s_i}^{j\rightarrow i},
\end{equation}
we can more succinctly write
\begin{equation}
    Z_G=\frac{\prod_{i=1}^N Z^i}{\prod_{ij\in E(G)}Z^{(ij)}}.
\end{equation}
From definition \eqref{eq:free entropy density given graph}, the expression \eqref{eq:free_entropy_graph} for the free entropy density as a function of the messages is obtained:
\begin{equation}\label{eq:free_entropy}
    \phi_G=\frac{1}{N}\sum_{i=1}^N \log Z^i -\frac{1}{N}\sum_{ij\in E(G)}\log Z^{ij}.
\end{equation}
The mean density of occupied nodes $\rho$ as a function of the message is given by
\begin{equation}
    \rho=\frac{\partial \phi}{\partial \mu}= \frac{1}{N}\sum_{i=1}^N \frac{\sum_{s_i}\sum_{\{s_k\}_{k\in\partial i}}s_ie^{\mu s_i} f(s_i, \{s_k\}_{k\in\partial i}) \prod_{k\in\partial i}\psi_{s_k, s_i}^{k\rightarrow i}}{Z^i}.
\end{equation}

\subsection{Belief Propagation Iterations on \texorpdfstring{$d$}{LG}-regular Graphs}\label{appendix:Belief propagation iterations on d-regular graphs}

We now consider the case of $d$-regular graphs. We will assume that the messages are locally equivalent, i.e. that they are the same independently of the nodes. This means that we reduce from $2|E(G)|$ messages to only one. The fixed point equation \eqref{eq:BP_equations_graph} then reduces to \eqref{eq:BP}, and we can iterate it:
\begin{equation}
      \tilde{\psi}_{s_i,s_j}(t+1)=
    \frac{1}{\tilde{Z}}\sum_{\{s_k\}_{k=1}^{d-1}} e^{\mu s_i} f(s_i, s_j, \{s_k\}_{k=1}^{d-1}) \prod_{k=1}^{d-1} \psi_{s_k,s_i}(t).
\end{equation}
where $t$ indicates the iteration and $\tilde{Z}$ is the normalization. In practice, the message is updated with a dampening $\epsilon \in [0, 1)$:
\begin{equation}
   \psi_{s_i,s_j}(t+1)=\epsilon \psi_{s_i,s_j}(t)+(1-\epsilon)\tilde{\psi}_{s_i,s_j}(t+1)
\end{equation}
The message can be represented by a $2\times 2$ matrix $\psi$ indexed by $s_i$ and $s_j$. We stop the iterations when the Frobenius norm of $\psi(t+1)-\psi(t)$ is smaller than a given tolerance (or after a maximum number of steps). We used a tolerance of $10^{-12}$ and a damping $\epsilon=0.8$.
The estimated free entropy \eqref{eq:free_entropy_graph} simplifies to
\begin{equation}\label{eq:explicit RS free entropy}
\phi_{\rm RS}=\log \left(\sum_{s_i, \{s_k\}_{k=1}^d} e^{\mu s_i} f(s_i,\{s_k\}_{k=1}^d) \prod_{k=1}^{d}\psi_{s_k,s_i}\right)
-\frac{d}{2}\log\left(\sum_{s_i, s_j}\psi_{s_i,s_j}\psi_{s_j,s_i}\right)
\end{equation}
which we call the replica symmetric free entropy density. The $d/2$ factor comes from the fact that a random d-regular graph of $N$ nodes has $\frac{Nd}{2}$ edges. This is the expression given in eq. \eqref{eq:RS_free_entropy} where we write $Z^i\rightarrow Z_\text{n}$ and $Z^{ij}\rightarrow Z_\text{e}$ to avoid referencing the graph.

\section{One-step Replica Symmetry Breaking Cavity Equation}\label{appendix: One-step replica symmetry breaking cavity equation}
In this annex we recover the 1RSB eq. \eqref{eq:1RSB}, the expression for the replicated free entropy density \eqref{eq:1RSB replicated free entropy density} and an explicit formula of the internal free entropy as function of the messages.

Suppose again that the graph is a tree. Let us write the self-consistent equation \eqref{eq:BP_equations} as
\begin{equation}
\begin{split}
    \psi^{i\rightarrow j}_{s_i,s_j}&=
    \frac{1}{Z^{i\rightarrow j}}\sum_{\{s_k\}_{k\in \partial i \setminus j}} e^{\mu s_i} f(s_i, s_j, \{s_k\}_{k\in\partial i\setminus j}) \prod_{k\in \partial i \setminus j} \psi_{s_k,s_i}^{k\rightarrow i}
    \\&=:\mathcal{F}(\{\mathbf{\psi}_{s_k,s_i}^{k\rightarrow i}\}_{k\in\partial i\setminus j}).
\end{split}
\end{equation}
If this is valid for every $s_i, s_j \in S$, then we write
\begin{equation}
    \psi^{i\rightarrow j}=\mathcal{F}\left(\{\psi^{k\rightarrow i}\}_{k\in \partial i\setminus j}\right)
\end{equation}
where a message $\psi^{a\rightarrow b}$ without the subscript indicates all the components of the message (e.g. in the form of a matrix).

Supposing that the internal free entropy density $\phi_{\rm int}(\tilde\psi)=\phi_{\text{Bethe}}(\tilde\psi)$ (i.e. that the Bethe-Peierls approximation holds within each cluster) and using the expression \eqref{eq:free_entropy} for $\phi_{\text{Bethe}}$, we can write the partition function of \eqref{eq:probability measure cluster}
\begin{equation}\label{eq:Z_mu}
\begin{split}
Z_\nu&=\sum_{\{\mathbf{\tilde\psi}\}}e^{Nm\phi_{\rm int}(\mathbf{\tilde\psi})}
\\&\stackrel{N\rightarrow \infty}{=}\int \prod_{ij\in E(G)}[d\psi^{i\rightarrow j}]e^{m\left(\sum_{i=1}^N\log Z^i - \sum_{ij\in E(G)}\log Z^{ij}\right)}
\\&\,\,\,\,\,\,\,\,\,\,\, \times \prod_{i=1}^N\prod_{j\in \partial i}\delta\left(\psi^{i\rightarrow j}-\mathcal{F}(\{\mathbf{\psi}^{k\rightarrow i}\}_{k\in\partial i\setminus j})\right)
\\&=\int \prod_{ij\in E(G)}[d\psi^{i\rightarrow j}]\prod_{i=1}^N \left(Z^i\right)^m\prod_{ij\in E(G)}\left(Z^{ij}\right)^{-m} 
 \prod_{i=1}^N\prod_{j\in \partial i}\delta\left(\psi^{i\rightarrow j}-\mathcal{F}(\{\mathbf{\psi}^{k\rightarrow i}\}_{k\in\partial i\setminus j})\right)
\end{split}
\end{equation}
where the integral is over all possible messages with the condition that they are normalized (i.e. that they are indeed summing to $1$) and $\delta(\cdot)$ is the Dirac delta. For notational purposes, we do not write the dependence of $Z^i$, $Z^{(ij)}$ to $\psi^{i\rightarrow j}$. We see that eq. \eqref{eq:Z_mu} is the partition function of the probability distribution
\begin{equation}\label{eq:probability_distribution_1RSB}
\begin{split}
    \tilde{P}\left(\{\psi^{i\rightarrow j}\}_{ij\in E(G)}\right)=&\frac{1}{Z_\mu} \prod_{i=1}^N \left(Z^i\right)^m\prod_{ij\in E(G)}\left(Z^{(ij)}\right)^{-m} 
    \\&\,\,\,\, \times \prod_{i=1}^N\prod_{j\in \partial i}\delta\left(\psi^{i\rightarrow j}-\mathcal{F}(\{\mathbf{\psi}^{k\rightarrow i}\}_{k\in\partial i\setminus j})\right).
\end{split}
\end{equation}
If the graph is a tree we can take the product over $ij$ such that the i's are never the same. Using the fact that $\delta(x)\delta(x)=\delta(x)$, equation \eqref{eq:probability_distribution_1RSB} becomes
\begin{equation}
        \tilde{P}\left(\{\psi^{i\rightarrow j}\}_{ij\in E(G)}\right)=\frac{1}{Z_\mu}\prod_{ij\in E(G)} \left(Z^i\right)^m \left(Z^{(ij)}\right)^{-m} \delta\left(\psi^{i\rightarrow j}-\mathcal{F}(\{\mathbf{\psi}^{k\rightarrow i}\}_{k\in\partial i\setminus j})\right).
\end{equation}
We can apply the cavity formalism on this probability distribution. Following the same procedure as in \ref{appendix:derivation BP on graphs}, the obtained self-consistent equation is
\begin{equation}\label{eq:BP_1RSB}
\begin{split}
        P^{i\rightarrow j}(\psi^{i\rightarrow j})=\frac{1}{\mathcal{Z}^{i\rightarrow j}}&\int \prod_{k\in\partial i\setminus j} [d \psi^{k\rightarrow i}]\left(\frac{Z^i}{Z^{ij}}\right)^m \delta\left(\psi^{i\rightarrow j}-\mathcal{F}(\{\mathbf{\psi}^{k\rightarrow i}\}_{k\in\partial i\setminus j})\right)
        \\ &\times\prod_{k\in\partial i\setminus j} [P^{k\rightarrow i}(\psi^{k\rightarrow i})]
\end{split}
\end{equation}
where $\mathcal{Z}^{i\rightarrow j}$ is the normalization. We will from here on write
\begin{equation}
\prod_{k\in\partial i\setminus j} [d \psi^{k\rightarrow i}]\times \prod_{k\in\partial i\setminus j} [P^{k\rightarrow i}(\psi^{k\rightarrow i})]=\prod_{k\in\partial i\setminus j} dP^{k\rightarrow i}.
\end{equation}
In the population dynamics algorithm (see Appendix~\ref{appendix: population dynamics}), we use the fact that $Z^i/Z^{ij}=Z^{i\rightarrow j}$. Indeed,
\begin{equation}
\begin{split}
\frac{Z^i}{Z^{ij}}&=\frac{\sum_{s_i}\sum_{\{s_k\}_{k\in\partial i}}e^{\mu s_i} f(s_i, \{s_k\}_{k\in\partial i}) \prod_{k\in\partial i}\psi_{s_k,s_i}^{k\rightarrow i}}{\sum_{s_i}\sum_{s_j}\psi_{s_i, s_j}^{i\rightarrow j} \psi_{s_j,s_i}^{j\rightarrow i}}
\\
&=\frac{\sum_{s_i}\sum_{\{s_k\}_{k\in\partial i}}e^{\mu s_i} f(s_i, \{s_k\}_{k\in\partial i}) \prod_{k\in\partial i}\psi_{s_k,s_i}^{k\rightarrow i}}{\sum_{s_i}\sum_{s_j} \left(\frac{1}{Z^{i\rightarrow j}}\sum_{\{s_k\}_{k\in \partial i \setminus j}} e^{\mu s_i} f(s_i, s_j, \{s_k\}_{k\in\partial i\setminus j}) \prod_{k\in \partial i \setminus j} \psi_{s_k,s_i}^{k\rightarrow i}\right)
\psi_{s_j,s_i}^{j\rightarrow i}}
\\
&=\frac{\sum_{s_i}\sum_{\{s_k\}_{k\in\partial i}}e^{\mu s_i} f(s_i, \{s_k\}_{k\in\partial i}) \prod_{k\in\partial i}\psi_{s_k,s_i}^{k\rightarrow i}}
{\frac{1}{Z^{i\rightarrow j}}\sum_{s_i}\sum_{s_j} \left(\sum_{\{s_k\}_{k\in \partial i \setminus j}} e^{\mu s_i} f(s_i, s_j, \{s_k\}_{k\in\partial i\setminus j}) \psi_{s_j,s_i}^{j\rightarrow i} \prod_{k\in \partial i \setminus j} \psi_{s_k,s_i}^{k\rightarrow i}\right)}
\\
&=Z^{i\rightarrow j}
\end{split}
\end{equation}
where we used the definitions \eqref{eq:Z^i}, \eqref{eq:Z^ij} of $Z^i$, $Z^{ij}$ in the first line and the BP eq. \eqref{eq:BP_equations} to go from the first to the second line.

The replicated free entropy density is obtained similarly as \eqref{eq:free_entropy}:
\begin{equation}\label{eq:Psi}
    \Psi(m)=\frac{1}{N}\sum_{i=1}^N\log \mathcal{Z}^i-\frac{1}{N}\sum_{ij\in E(G)}\log \mathcal{Z}^{ij}
\end{equation}
with
\begin{equation}
    \mathcal{Z}^i:=\int \prod_{k\in\partial i} dP^{k\rightarrow i} \left(Z^{i}\right)^m,
\end{equation}
\begin{equation}
    \mathcal{Z}^{ij}:=\int dP^{i\rightarrow j}dP^{j\rightarrow i} \left(Z^{ij}\right)^m.
\end{equation}

To obtain the free entropy density of a cluster, we use
\begin{equation}\label{eq:free_entropy_of_cluster}
\begin{split}
    \phi_{\rm int}=\frac{\partial \Psi(m)}{\partial m} = & \frac{1}{N}\sum_{i=1}^N  \frac{1}{\mathcal{Z}^i} \int \prod_{k\in\partial i} dP^{k\rightarrow i} \log( Z^i)\left(Z^{i}\right)^m
    \\ & - \frac{1}{N}\sum_{ij\in E(G)}\frac{1}{\mathcal{Z}^{ij}}\int dP^{i\rightarrow j}dP^{j\rightarrow i} \log(Z^{ij}) \left(Z^{ij}\right)^m.
\end{split}
\end{equation}
The density is given by
\begin{equation}\label{eq:rho_1RSB}
\begin{split}
    \rho&=\frac{\partial \phi}{\partial{\mu}}=\frac{1}{N}\sum_{i=1}^N\int\prod_{k\in\partial i}d P^{k\rightarrow i}\frac{\partial}{\partial \mu}\left(\frac{\log(Z^i) (Z^i)^m}{\mathcal{Z}^i}\right)
    \\&=\frac{1}{N}\sum_{i=1}^N\int\prod_{k\in\partial i}d P^{k\rightarrow i}\left( \frac{\partial_{\mu}Z^i}{Z^i}\frac{(Z^i)^m}{\mathcal{Z}^i}+\frac{\log(Z^i)^2 (Z^i)^m \partial_\mu Z^i}{\mathcal{Z}^i}-\frac{\log(Z^i) (Z^i)^m}{(\mathcal{Z}^i)^2}\partial_\mu \mathcal{Z}^i\right)
    \\&=\frac{1}{N}\sum_{i=1}^N\int\prod_{k\in\partial i}d P^{k\rightarrow i}\frac{(Z^i)^{m-1}\partial_\mu Z^i}{\mathcal{Z}^i}
\end{split}
\end{equation}
where we notice that $Z^{ij}$, $\mathcal{Z}^{ij}$ do not depend on the Lagrange multiplier $\mu$ and the two last terms of the second line cancel each other out.

Moving to d-regular graphs, we will suppose again that the messages are locally equivalent. Thus, there is no more dependence on the graph and we use the RS free entropy density~\eqref{eq:RS_free_entropy} instead of the Bethe free entropy density~\eqref{eq:free_entropy_graph}. Eq. \eqref{eq:BP_1RSB} becomes eq. \eqref{eq:1RSB} using the notation $dP^{a\rightarrow b}\rightarrow dP(\psi)$. Similarly, the replicated free entropy density \eqref{eq:Psi} becomes \eqref{eq:1RSB replicated free entropy density} following the same procedure as for eq. \eqref{eq:explicit RS free entropy} with $\mathcal{Z}^i\rightarrow \mathcal{Z}_n$ and $\mathcal{Z}^{ij}\rightarrow \mathcal{Z}_e$. The internal free entropy density \eqref{eq:free_entropy_of_cluster} becomes
\begin{equation}
      \phi_{\rm int}=  \frac{1}{\mathcal{Z}_n} \int \prod_{k=1,...,d} dP(\psi^{(k)}) \log( Z_\text{n})\left(Z_\text{n}\right)^m- \frac{d}{2}\frac{1}{\mathcal{Z}_e} \int 
 dP(\psi^{(1)}) dP(\psi^{(2)}) \log(Z_\text{e}) \left(Z_\text{e})\right)^m .
\end{equation}
The complexity can be obtained from \eqref{eq:Laplace replicated free entropy density} and is given by
\begin{equation}
    \Sigma(\phi_{\rm int})=\Psi(m)-m\phi_{\rm int}.
\end{equation}
The density \eqref{eq:rho_1RSB} becomes
\begin{equation}
    \rho=\int\prod_{k=1,...,d}d P(\psi^{(k)})\frac{(Z_\text{n})^{m-1}\partial_\mu Z_\text{n}}{\mathcal{Z}_n} .
\end{equation}

\section{Population Dynamics}\label{appendix: population dynamics}
The pseudocode for population dynamics algorithms can be found for instance in \cite{zdeborova_statistical_2008} or \cite{Information_physics_and_computations}. Below we describe it informally:
\begin{enumerate}
    \item Initialize a list of $M$ messages $\psi$. This list is called the \textit{population}
    \item Choose uniformly at random $L=\lceil(1-\epsilon)M\rceil$ messages from the population. We will call them the \textit{subpopulation}. $\epsilon$ is the \textit{damping parameter} and plays a similar role as in the BP algorithm.
    \item Update each message of the subpopulation by doing one BP iterations. The BP iteration is done by sampling $d-1$ neighbors uniformly at random from the population for each message of the subpopulation.
    \item Compute the normalization $Z_l$ for each message of the updated subpopulation.
    \item Sample $L$ messages from the updated subpopulation from the distribution $P(l)=\frac{Z_l^m}{\sum_{i=1}^L Z_i^m}$.
    \item Replace the subpopulation from the sampled messages from step 5.
    \item Repeat from step 2.
\end{enumerate}
The algorithm is stopped when the obtained distribution stabilizes (this is checked by comparing distances between histograms of the population), or after a maximal number of iterations. In the case were the algorithm is used to determine the stability of the RS solution, the messages of the population are initialize on noisy copies of the RS solutions. To approximate the solution of the 1RSB equation, the messages of the population are initialized on \textit{hard-fields}, i.e. one component of the message is $1$ and the others are $0$, as prescribed by the reconstruction on trees presented in Section \ref{sec:Structure of the space of solutions}.

The observables are then sampled from the population. This is done by sampling uniformly at random $K$ messages from the population and computing the observables by approximating the integrals \eqref{eq:Psi} and \eqref{eq:free_entropy_of_cluster}. This is repeated many times, and the average is then taken. Since the population can still fluctuate after many iterations, the computation of the observables can be done during the iterations of the population dynamics, after a stabilization of the population. However, to avoid correlations, sampling should not be done at each iteration. The average over the observables at different iterations is then taken, and the standard deviation is used to indicate the error.

This task was vectorized and implemented using pytorch \cite{pytorch} to allow for large populations (we use $M=10^7$) to be run in acceptable time. We use $\epsilon=0.8$ and we consider that the populations stabilize after $8000$ iterations or when the distribution of the messages in the population ceases to change significantly. The sampling of the observables is then done with $K=2\cdot10^8$ every $50$ iteration for a total of $2000$ iterations.

For the rules that are locked (when $\phi_{\rm int}=0)$, the numerical result is not exactly zero due to the fact that a variable can be frozen but the message is not a hard-field (e.g. for $\psi_{0,0}=\psi_{0,1}=0.5$). This can lead to numerical errors due to the estimation of the integrals \eqref{eq:cal Z_n}, \eqref{eq:cal Z_e} and the fluctuations of the population. To circumvent this problem, we check if each message of the population implies the value of a node, and put the internal free entropy to~$0$ in Tables \ref{tbl:classification_with_homogeneous} and \ref{tbl:classification_without_homogeneous} if this is the case.

\section{Local Entropy}\label{appendix:local entropy}

Following the procedure of Appendix \ref{appendix:derivation BP on graphs} with the partition function given by \eqref{eq:Partition function local entropy}, we obtain the following BP equations on a graph of size $N$:
\begin{equation}\label{eq:BP equations local entropy}
    \psi^{i\rightarrow j}_{s_i, s_j}=
    \frac{1}{Z^{i\rightarrow j}}\sum_{\{s_k\}_{k\in \partial i \setminus j}} e^{\kappa \mathds{1}(s'_i\neq s_i)} f(s_i, s_j, \{s_k\}_{k\in\partial i\setminus j}) \prod_{k\in \partial i \setminus j} \psi_{s_k, s_i}^{k\rightarrow i}.
\end{equation}
where we did not write the dependence to the given configuration $\mathbf{s}'$, and suppose that this configuration is a solution so as to avoid an additional indicator function. $Z^{i\rightarrow j}$ is the normalization such that $\sum_{s_i} \sum_{s_j} \psi^{i\rightarrow j}_{s_i, s_j} = 1$. 

The free entropy density is given by
\begin{equation}\label{eq:free_entropy_LE}
    \phi_{LE}=\frac{1}{N}\sum_{i=1}^N \log Z^i -\frac{1}{N}\sum_{(ij)\in E(G)}\log Z^{ij}.
\end{equation}
where
\begin{equation}
    Z^i=\sum_{s_i}\sum_{\{s_k\}_{k\in\partial i}}e^{\kappa \mathds{1}(s'_i\neq s_i)} f(s_i, \{s_k\}_{k\in\partial i})\prod_{k\in\partial i}\psi_{s_k,s_i}^{k\rightarrow i}
\end{equation}
\begin{equation}
    Z^{ij}=\sum_{s_i}\sum_{s_j}\psi_{s_i, s_j}^{i\rightarrow j} \psi_{s_j,s_i}^{j\rightarrow i}.
\end{equation}

The (normalized) distance $d(\mathbf{s}, \mathbf{s}')$ between  two configurations $\mathbf{s}$ and $\mathbf{s}'$ is defined as
\begin{equation}
    d(\mathbf{s}, \mathbf{s}')=\frac{\sum_{i=1}^N \mathds{1}(s_i\neq s'_i)}{N},
\end{equation}
where the indicator function $\mathds{1}$ is $1$ if the argument is true, and $0$ otherwise. In the large $N$ limit, the average distance is very peaked and is equal to the distance. The distance to $\mathbf{s}'$ as function of the messages is given by
\begin{equation}\label{eq:distance}
    d=\frac{\partial \phi_{LE}}{\partial \kappa}= \frac{1}{N}\sum_{i=1}^N \frac{\sum_{s_i}\sum_{\{s_k\}_{k\in\partial i}}e^{\kappa \mathds{1}(s'_i\neq s_i)} f(s_i, \{s_k\}_{k\in\partial i}) \prod_{k\in\partial i}\psi_{s_{(ki)}}^{k\rightarrow i}}{Z^i}.
\end{equation}

Let $\mathcal{N}(\mathbf{s}', d)$ be the exponential leading term of the number of solutions at distance $d$ of $\mathbf{s}'$. The local entropy density $s_{LE}(\mathbf{s},d)$ is defined as $\mathcal{N}(\mathbf{s}', d)=e^{N s_{LE}(\mathbf{s}',d)}$. Writing 
\begin{equation}
    e^{N\phi_{LE}}=\sum_{d=i/N, i=0,\hdots, N}e^{\kappa N d (\mathbf{s}', d)} \mathcal{N}(\mathbf{s}', d)
\end{equation}
and applying the Laplace method we obtain
\begin{equation}\label{eq:local entropy}
    s_{LE}=\phi_{LE}-\kappa d.
\end{equation}
This standard reasoning of statistical physics is also used to obtain eq. \eqref{eq:Laplace replicated free entropy density}.

The Legendre transform only gives the concave envelope of the local entropy. To obtain non-concave behavior, we can fix the distance $d^*$ and adapt $\kappa$ to obtain this distance. In practice, at every $t_\kappa$ iterations of the BP eq. \eqref{eq:BP equations local entropy} we update $\kappa\leftarrow \kappa+\gamma(d^*-d)$, where $d$ is the distance computed from \eqref{eq:distance} at the current BP messages. $\gamma$ is a factor determining how much $\kappa$ is changed. We use $\gamma=5$ and $t_\kappa=1$ to obtain Figures \ref{fig:LE}. The pseudocode is given in Algorithm \ref{algo:BP local entropy at fixed distance}. After convergence or after $T$ steps, the free local entropy, the distance and the local entropy can be computed from the messages using eqs. \eqref{eq:free_entropy_LE}, \eqref{eq:distance} and \eqref{eq:local entropy}.

\begin{algorithm}[ht]
\caption{BP-LOCAL ENTROPY $(G, \mathbf{s}',  d^*, T, \gamma, t_{\kappa}, \epsilon, \epsilon_{BP}, \epsilon_d)$}
\label{algo:BP local entropy at fixed distance}
\begin{algorithmic}[1]
    \STATE Initialize the messages $\psi_{s_i, s_j}^{i\rightarrow j}$ randomly on the factor graph of $G$ and normalize them, we denote $\psi^{i\rightarrow j}=\begin{pmatrix}\psi^{i\rightarrow j}_{0,0} & \psi^{i\rightarrow j}_{0,1} \\ \psi^{i\rightarrow j}_{1,0} & \psi^{i\rightarrow j}_{1,1}\end{pmatrix}$ a $2\times 2$ matrix and $\psi$ the list of all the messages.
    \STATE $\psi^{i\rightarrow j}_{old} \leftarrow \psi^{i\rightarrow j}$
    \STATE $\kappa\leftarrow 0$
    \STATE $t \leftarrow 0$
    \WHILE{$\frac{1}{2|E|}\sum_{(i\rightarrow j)} ||\psi^{i\rightarrow j}-\psi^{i\rightarrow j}_{old}||_F>\epsilon_{BP}$ and $|d(\psi)-d^*|>\epsilon_d$ and $t \leq T$}
        \STATE $\psi^{i\rightarrow j}_{old}\leftarrow \psi^{i\rightarrow j}$
        \IF{$t_\kappa$ divides $t$}
            \STATE $\kappa \leftarrow \kappa+\gamma(d^*-d(\psi))$
        \ENDIF
        \STATE Update all the messages $\psi_{s_i, s_j}^{i\rightarrow j}$ according to \eqref{eq:BP equations local entropy} with dampening $\epsilon$.
        \STATE $t \leftarrow t+1$
    \ENDWHILE
    \RETURN $\psi$, $\kappa$.
\end{algorithmic}
\end{algorithm}

\section{Equivalence between the Occupation Problems and the \texttt{+-} Rules}\label{appendix:occupation problems}
The occupation problems studied in \cite{zdeborova_constraint_2008} can be seen as the rules containing only \texttt{+} and \texttt{-} (the rules does not depend on the state of the middle node). It was found that, even though the graphical model is not the same, the entropy matches.

Fixing $L=K=d$, \cite{zdeborova_constraint_2008} gives the following BP fixed point equation:
\begin{equation}\label{eq:BP_Zdeborova_occupation}
\begin{split}
    \psi_0&=\frac{1}{Z_{reg}}\sum_{r=0}^{d-1}A_r \binom{d-1}{r}\psi_1^{(d-1)r}\psi_0^{(d-1)(d-1-r)} \\
    \psi_1&=\frac{1}{Z_{reg}}\sum_{r=0}^{d-1}A_{r+1} \binom{d-1}{r}\psi_1^{(d-1)r}\psi_0^{(d-1)(d-1-r)}
\end{split}
\end{equation}
where $A_r$ is $1$ if the symbol at position $r+1$ of the rule is a \texttt{+} and $0$ otherwise.

Our eq. \eqref{eq:BP} simplifies in this case to 
\begin{equation}
\begin{split}
    \psi_{0,0}&=\frac{1}{Z_\psi}\sum_{r=0}^{d-1}A_r \binom{d-1}{r}\psi_{0,0}^{d-1-r}\psi_{1,0}^r\\
    \psi_{1,0}&=\frac{1}{Z_\psi}\sum_{r=0}^{d-1}A_r \binom{d-1}{r}\psi_{0,1}^{d-1-r}\psi_{1,1}^r\\
    \psi_{0,1}&=\frac{1}{Z_\psi}\sum_{r=0}^{d-1}A_{r+1} \binom{d-1}{r}\psi_{0,0}^{d-1-r}\psi_{1,0}^r\\
    \psi_{1,1}&=\frac{1}{Z_\psi}\sum_{r=0}^{d-1}A_{r+1} \binom{d-1}{r}\psi_{0,1}^{d-1-r}\psi_{1,1}^r.
\end{split}
\end{equation}

$\psi_{s_i}$ indicates the probability that a node takes the value $s_i$ when we have a cavity. We verified numerically that $\psi_{s_i}=\psi_{0, s_i}+\psi_{1, s_i}$ on all the occupation rules for $d=3,4$.

Summing $\psi_{0,0}+\psi_{1,0}$ and $\psi_{0,1}+\psi_{1,1}$ we obtain
\begin{equation}\label{eq:B_fixed_point}
\begin{split}
    \psi_{0,0}+\psi_{1,0}&=\sum_{r=0}^{d-1}A_r \binom{d-1}{r}\frac{1}{Z_\psi}\overbrace{(\psi_{0,0}^{d-1-r}\psi_{1,0}^r+\psi_{0,1}^{d-1-r}\psi_{1,1}^r)}^{:= B(r)}\\
    \psi_{0,1}+\psi_{1,1}&=\sum_{r=0}^{d-1}A_{r+1}\binom{d-1}{r}\frac{1}{Z_\psi}(\psi_{0,0}^{d-1-r}\psi_{1,0}^r+\psi_{0,1}^{d-1-r}\psi_{1,1}^r).
\end{split}
\end{equation}

Let $\psi_0=\psi_{0,0}+\psi_{1,0}$ and $\psi_1=\psi_{0,1}+\psi_{1,1}$. The BP equations \eqref{eq:BP_Zdeborova_occupation} become
\begin{equation}\label{eq:C_fixed_point}
\begin{split}
\psi_0=\sum_{r=0}^{d-1}A_r \binom{d-1}{r}
\frac{1}{Z_{reg}}\overbrace{ \left(\sum_{r'=0}^{d-1}\binom{d-1}{r'} \psi_{0,1}^{d-1-r'}\psi_{1,1}^{r'}\right)^r
\left(\sum_{r''=0}^{d-1}\binom{d-1}{r''} \psi_{0,0}^{d-1-r''}\psi_{1,0}^{r''}\right)^{d-1-r}}^{:=C(r)}
\\
\psi_1=\sum_{r=0}^{d-1}A_{r+1} \binom{d-1}{r}
\frac{1}{Z_{reg}} \left(\sum_{r'=0}^{d-1}\binom{d-1}{r'} \psi_{0,1}^{d-1-r'}\psi_{1,1}^{r'}\right)^r
\left(\sum_{r''=0}^{d-1}\binom{d-1}{r''} \psi_{0,0}^{d-1-r''}\psi_{1,0}^{r''}\right)^{d-1-r}
\end{split}
\end{equation}
where we used the binomial expansion of $(\psi_{0,0}+\psi_{1,0})^{d-1}$  and $(\psi_{0,1}+\psi_{1,1})^{d-1}$. 

We would like to show that the BP equations are equivalent, which is the case if $B(r) Z_{reg}=C(r) Z_\psi$ for every $r$. We verified numerically that the found fixed point satisfies this equation for each occupation problem for $d=3,4$. Additionally, inspecting the numerical values of the fixed point we notice that
\begin{equation}
\psi_{0,1}\psi_{1,0}=\psi_{0,0}\psi_{1,1}
\end{equation}
\begin{equation}
\psi_{1,0}^{d-1}=\psi_{1,1}^{d-2}\psi_{0,1}.
\end{equation}
However, we do not know how to justify these relations. From these equations, we obtain the substitutions
\begin{equation}
\psi_{0,1}=\frac{\psi_{1,0}^{d-1}}{\psi_{1,1}^{d-2}}
\end{equation}
and 
\begin{equation}
\psi_{0,0}=\frac{\psi_{0,1}\psi_{1,0}}{\psi_{1,1}}=\frac{\psi_{1,0}^d}{\psi_{1,1}^{d-1}}.
\end{equation}
Using these substitutions we obtain analytically the desired relation $ B(r) Z_{reg}=C(r)Z_\psi \,\forall r$ for $d=3,4,5,6,7$.

\end{document}